\documentclass[preprint,5p,10pt]{elsarticle}

\usepackage[ruled,linesnumbered,vlined]{algorithm2e}
\usepackage[ruled]{algorithm2e}
\SetAlFnt{\footnotesize}
\SetAlCapFnt{\footnotesize}
\SetAlCapNameFnt{\footnotesize}
\usepackage[caption=false,font=footnotesize]{subfig}
\usepackage{caption}
\usepackage{subcaption}
\usepackage{url}
\usepackage[table]{xcolor}
\usepackage{nameref}
\usepackage{paralist}
\usepackage{multicol}
\usepackage{multirow}
\usepackage{graphicx}
\usepackage[fleqn]{amsmath}
\usepackage{amsfonts}
\usepackage{tcolorbox}
\usepackage{soul}
\usepackage{flushend}
\usepackage{capt-of}
\usepackage{lmodern}
\usepackage{booktabs}

\SetCommentSty{mycommfont}
\let\oldnl\nl
\newcommand{\nonl}{\renewcommand{\nl}{\let\nl\oldnl}}

\DeclareMathOperator*{\maximize}{maximize}
\DeclareMathOperator*{\minimize}{minimize}
\usepackage{mathtools}
\DeclarePairedDelimiter\ceil{\lceil}{\rceil}

\journal{Computer Networks}

\begin{document}

\begin{frontmatter}

\title{Toward Scalable VR-Cloud Gaming: An Attention-aware Adaptive Resource Allocation Framework for 6G Networks}

\author[label1]{Gabriel M. Almeida} \ead{gabrielmatheus@inf.ufg.br}
\author[label1]{João Paulo Esper}
\author[label2]{Cleverson Nahum}
\author[label2]{Aldebaro Klautau}
\author[label1]{Kleber Vieira Cardoso}

\cortext[cor1]{Corresponding author.}

\address[label1]{Instituto de Informática (INF), Universidade Federal de Goiás (UFG), Goiânia, Brazil - 74690-900}
\address[label2]{Universidade Federal do Pará (UFPA), Belém, Brazil - 66075-750}

\begin{abstract}
Virtual Reality Cloud Gaming (VR-CG) represents a demanding class of immersive applications, requiring high bandwidth, ultra-low latency, and intelligent resource management to ensure optimal user experience. In this paper, we propose a scalable and QoE-aware multi-stage optimization framework for resource allocation in VR-CG over 6G networks. Our solution decomposes the joint resource allocation problem into three interdependent stages: (i) user association and communication resource allocation; (ii) VR-CG game engine placement with adaptive multipath routing; and (iii) attention-aware scheduling and wireless resource allocation based on motion-to-photon latency.
For each stage, we design specialized heuristic algorithms that achieve near-optimal performance while significantly reducing computational time. We introduce a novel user-centric QoE model based on visual attention to virtual objects, guiding adaptive resolution and frame rate selection.
A dataset-driven evaluation demonstrates that, when compared against state-of-the-art approaches, our framework improves QoE by up to 50\%, reduces communication resource usage by 75\%, and achieves up to 35\% cost savings, while maintaining an average optimality gap of 5\%. Our proposed heuristics solve large-scale scenarios in under 0.1 seconds, highlighting their potential for real-time deployment in next-generation mobile networks.
\end{abstract}


\end{frontmatter}

\section{Introduction}
\label{sec:intro}

The sixth generation (6G) of cellular networks is expected to redefine wireless connectivity, providing unprecedented support for densely populated environments and enabling a new wave of applications that demand high data rates and ultra-low latency communication~\cite{Saad:2020}. To meet this unprecedented connectivity demand, both industry and academia have proposed a range of innovative technologies aimed at enhancing network capacity, reliability, and adaptability. Among the most promising enablers are Terahertz (THz) communications~\cite{Shafie:23}, Computing and Network Convergence (CNC) architectures~\cite{Tang:2024}, and Semantic Communication (SC) systems~\cite{Xia:2023}.

At the core of these technologies lie the challenging use cases envisioned for 6G networks~\cite{Giordani:2020}. Building upon the achievements of 5G, which enabled reliable high-resolution video streaming over mobile networks, 6G aims to extend this capability to support immersive applications. This emerging class of applications is grounded in the concept of Virtual Reality (VR), where users interact with virtual environments through Head-Mounted Displays (HMDs), such as VR headsets. These devices demand ultra-high-resolution content for both eyes and high frame rates to ensure seamless and realistic user experiences. However, beyond delivering higher resolution images, immersive applications are designed to enrich user experiences through interactive virtual environments, where users can manipulate virtual objects and dynamically alter the environment in response to their actions.

In this way, a key challenge arises concerning Head-Mounted Display (HMD) devices. Due to their wearable nature, HMDs must be compact, lightweight, and energy efficient. However, immersive applications require the efficient rendering of ultra-high-resolution images for both eyes, along with precise motion tracking to react to users actions manipulating the virtual environment. Meeting these stringent requirements demands high-performance processing and rendering capabilities within the HMD, which is challenging given its inherent constraints in terms of size, weight, temperature, and battery life.

As a response to this challenge, the concept of Virtual Reality Cloud Gaming (VR-CG) is proposed as one of the most prominent and challenging use cases of immersive applications in future networks \cite{TR38838:2022}. Unlike traditional VR streaming, VR-CG introduces a highly interactive dimension, where users can manipulate virtual objects through controller inputs and body movements, enabling them to move, hold, throw, or rotate virtual objects. This high level of interactivity introduces unique technical challenges for both communication and rendering systems \cite{Xia:2023,TR38838:2022,huawei:2018,Ruan:2021}.

Due to the highly dynamic, interaction between users and virtual objects, VR-CG applications impose two additional and unique requirements on system design:
(i) virtual objects are subject to continuous user manipulation, necessitating fine-grained object discretization and efficient processing to maintain visual coherence and responsiveness;
(ii) user interaction vary significantly across individuals depending on factors such as gaming experience, play style, and real-time decisions, which directly affect resource demands in both computing and communication domains.
While traditional VR applications benefit from well-established Quality of Experience (QoE) models, based on image quality and network Key Performance Indicators (KPIs), such models are insufficient for VR-CG \cite{Almeida:2025}. The critical role of user interactivity in VR-CG fundamentally alters the nature of user experience, dynamically influencing the rendering system, resource demands, and traffic shape. Due to the highly variable user behavior in VR-CG, universal QoE models fail to capture individual differences in users' interactions. Consequently, there is a need for personalized QoE models that adapt to each user’s unique engagement with the virtual environment \cite{Du:2024}.

Furthermore, traditional resource allocation strategies, which typically optimize rendering and network Quality of Service (QoS) metrics, are poorly suited to the dynamic, user-driven nature of VR-CG~\cite{Yang:2023}. As user actions continuously reshape computing and network resource demands (e.g., changes in scene complexity, frame rendering rates, and latency constraints), an effective solution requires an adaptive, user-centric resource allocation framework that responds to evolving user behavior in real time.

To address these challenges, we consider the combination of HMD sensory info and Computer Vision (CV) techniques to enable personalized, user-centric resource allocation in VR-CG applications~\cite{Du:2023}. By using AI/ML-based object segmentation models alongside eye-tracking data from HMD sensors, it is possible to map user attention to individual virtual objects within the FoV. This enables the extraction of a personalized KPI that reflects the dynamic interaction between the user and the virtual environment, allowing for fine-grained, attention-aware resource allocation~\cite{Du:2024}. Objects that receive greater user attention can be rendered at higher resolutions and with greater computing priority, while less attended elements can be optimized accordingly. This approach not only enhances QoE, but also improves resource usage in 6G networks.

In this work, we address the formulation of the VR-CG resource allocation problem in 6G network systems. We jointly optimize user QoE and resource allocation, introducing a novel user attention-driven QoE model for VR-CG applications. To enable practical scalability, we further propose a multi-stage optimization framework that decomposes the NP-hard joint problem into subproblems. The contributions of this paper are summarized as follows:

\begin{itemize}
\item \textbf{User-centric QoE model for VR-CG applications --} We propose a novel QoE model that leverages user attention to virtual objects, enabling personalized and dynamic QoE estimation based on user interaction with the virtual environment.
\item \textbf{Adaptive network resource allocation --} By integrating computer vision-based object segmentation with AI/ML-driven attention prediction, our framework continuously updates attention KPIs to guide adaptive resource allocation, ensuring that network and computing resources align with user priorities.
\item \textbf{Multi-stage optimization framework for 6G networks --} We introduce a three-stage optimization framework that decomposes the joint VR-CG resource allocation problem into tractable subproblems, reducing computational complexity and supporting efficient exploration of solution spaces.
\item \textbf{Scalable resource allocation --} We propose a set of heuristics that enable fast and scalable decision-making for VR-CG resource allocation. These methods introduce novel strategies to jointly account for QoE requirements, network conditions, and computational constraints, offering practical mechanisms that complement the optimization model and support efficient operation in complex 6G scenarios.
\item \textbf{Performance evaluation --} Among extensive simulations using real-world datasets, we evaluate the effectiveness of our solutions in improving personalized QoE and optimizing both computing and communication resources. Our proposed heuristic algorithms achieves up to 35\% cost and 75\% communication resource savings over state-of-the-art solutions.
\item \textbf{Reproducibility --} The code implementing both the optimization model and the proposed heuristics, along with the datasets and details of the simulation environment, is publicly available, enabling full replication of our experiments\footnote{https://github.com/LABORA-INF-UFG/paper-GJLK-2025}.
\end{itemize}

The remainder of this paper is organized as follows. Section~\ref{sec:related-work} reviews the related work and positions our work within the existing literature. Section~\ref{sec:VR_CG_network} defines the VR-CG system, detailing the application architecture, network environment, and attention-aware resource allocation mechanism. In Section~\ref{sec:system_model}, we describe the system model adopted in the problem formulation and formulate the VR-CG resource allocation problem. Section~\ref{sec:VR-CG_solutions} presents our heuristic algorithms designed to address the proposed formulation, taking into account the practical constraints of real-world networks. Section~\ref{sec:evaluation} provides the performance evaluation of our proposals and compares them against existing state-of-the-art solutions from the literature. Finally, Section~\ref{sec:conclusions} presents our final consideration and future work.
\section{Related Work}
\label{sec:related-work}

In this section, we present a brief review of the existing literature on VR-CG. We present the works considering three categories: QoE modeling for VR applications, semantic communication video transmission, and VR-CG resource allocation problem. Each category addresses a distinct aspect of system design and collectively contributes to a comprehensive understanding of the challenges and potential solutions in VR-CG systems.

\subsection{QoE modeling for VR applications}

This section introduces works that present QoE modeling advances to VR applications, considering objective metrics, such as video quality and connectivity metrics, and subjective metrics, such as users' attention and physiological metrics.

A comprehensive evaluation of \textit{Air Light VR}, a cloud gaming platform for VR, is presented in \cite{Li:2020}. By analyzing system-level QoS metrics and user-reported QoE, the authors show that packet loss and bandwidth limitations degrade user experience more than latency. Moderate latency variations had limited influence on perceived QoE, challenging traditional latency assumptions in VR service delivery. Nonetheless, this work lacks a formal problem formulation and relies exclusively on small-scale evaluations, limiting its scalability to more complex scenarios.

Aiming to quantify QoE under diverse conditions, in \cite{Lee:2024} the authors introduce a testbed that simulates varying network conditions for cloud-VR gaming. The authors conduct opinion-based studies to evaluate the influence of video encoding parameters, game genres, and network variability in users experience. Their results provide fine-grained insights into the interplay between technical parameters and subjective perception. However, the experimental setup relies solely on Wi-Fi Local Area Network (LAN) connectivity between HMDs and local server, failing to account for challenges inherent to cellular networks.

To support simulation-based studies, the authors of \cite{Zhao:2021}, presents a statistical modeling of VR-CG video traffic. Using commercial VR equipment and real-world games, the study characterizes frame size, inter-arrival time, and latency under various encoding schemes. The traffic models can support the design of more realistic evaluation scenarios for VR applications. However, the authors do not explore a formulation for VR-CG resource allocation.

In \cite{Warsinke:2024}, the authors explore how different types of network degradation affect users' QoE by streaming two VR games over \textit{Meta Air Link}. Using a within-subjects experiment, the authors demonstrate that packet loss is more detrimental to user comfort than latency, increasing cybersickness and decreasing the feeling of presence. While the findings are insightful, the experiments are limited to Wi-Fi networks and do not extend to cellular environments, which are essential for scalable and mobile VR deployment.

Finally, in \cite{Begue:2023,Vijayakumar:2024} the authors explore the use of physiological signals to assess user QoE across different video quality settings. Leveraging the dataset introduced in \cite{Perrin:2015}, the authors implement QoE prediction mechanisms based on features extracted from electroencephalogram and electrocardiogram data. While their findings demonstrate the potential of physiological data for enhancing QoE assessment, these works do not provide a formal QoE model. Instead, they focus on identifying the most QoE-relevant features within the dataset, offering a foundation for future modeling efforts in immersive application scenarios.

\subsection{Semantic video transmission}

Several works have proposed semantic-aware video encoding techniques to reduce bandwidth consumption in VR-CG while maintaining high perceptual quality. The authors of \cite{Zou:2021} introduce an object-aware encoding mechanism that combines scene information from the game engine with user eye-tracking data. By prioritizing video quality in areas where the user is focused and compressing peripheral regions more aggressively, the approach effectively reduces transmission requirements. However, the method assumes full integration between the game engine and encoding process, which may limit its general applicability to heterogeneous systems or games with limited engine access.

Following a similar idea, the authors of \cite{Song:2023} propose a gaze-based adaptive encoding strategy that classifies users by profile and experience level. This enables real-time identification of regions of interest based on gaze tracking and personalized video quality adjustments. While this user-based strategy offers an efficient use of resources and improved user experience, the system design is evaluated primarily in controlled settings and lacks scalability validation in dynamic or dense user environments.

In \cite{Xia:2023}, the authors explore semantic communication as a paradigm shift for future wireless systems, advocating for a transition from conventional bit-level communication to meaning-level transmission. The study underscores the importance of AI/ML in enabling this transformation and reviews recent developments in deep learning-based end-to-end systems. Although the work provides valuable theoretical insights and outlines critical research challenges, it lacks concrete application scenarios or validation specific to VR-CG or immersive applications.

Lastly, in  \cite{Vibhaalakshmi:2024} Gemino is introduced, a neural video compression framework tailored for real-time communication under extremely low bitrate conditions. By using a high-frequency super-resolution pipeline, Gemino achieves high-quality video reconstruction from low-resolution inputs. Despite its impressive results, demonstrating superior performance over traditional codecs, the system is designed specifically for video conferencing applications.

\subsection{VR-CG resource allocation}

Several recent studies have addressed different aspects of the resource allocation problem in immersive applications. One approach is presented in \cite{Wang:2018}, where they investigate the placement of VR applications on edge computing infrastructure by jointly optimizing operational cost and QoS metrics. The authors formulate the problem as an unconstrained combinatorial optimization task and propose a graph-cut-based iterative algorithm to determine the optimal placement of VR applications. While effective in reducing operational costs and improving QoS using real wireless network traces, the study focuses solely on application placement and does not account for the communication resource allocation in cellular networks.

However, in \cite{Vaidya:2023}, the authors address network resource optimization through predictive modeling of VR-CG video frame sizes, aiming to improve overall transmission efficiency. The authors evaluate AI/ML models on a testbed and propose an online transfer learning strategy to generalize predictions across network conditions. Although their model can capture the dynamic nature of network and application layers, it lacks a formal optimization formulation for resource allocation in cellular environments and is not evaluated in multi-user or large-scale network settings.

A broader solution is proposed in \cite{Zhang:2017}, which targets the scalability and responsiveness demands of VR Massively Multiplayer Online Games (VR-MMOGs). The authors present an edge-cloud architecture where rendering and view updates are processed at the edge, and global state management is centralized. A Markov Decision Process (MDP) is used to dynamically migrate services to co-locate them with users, reducing latency and migration overhead. Despite its scalability features, the architecture is not evaluated in the context of cellular infrastructure, limiting its applicability to real-world mobile deployments.

A more recent work, in \cite{Du_2023}, explicitly considers cellular networks and proposes an attention-aware resource allocation model for immersive \textit{Metaverse}\footnote{https://www.meta.com/metaverse/} applications. The formulation aims to maximize QoE using an object-aware transmission strategy based on user interaction and attention levels. Although this represents a significant step toward user-centric and AI-driven resource allocation, the model assumes a single Base Station (BS) environment, which restricts its scalability and diverges from the realistic multi-BS deployments expected in future 6G networks.

Our recent work, in \cite{Almeida:2024}, advances this line of research by formulating a joint optimization framework for the VR-CG resource allocation problem, covering user association, transmission resource allocation, and application placement over a 6G network infrastructure. We demonstrate that the formualted problem is NP-hard and present a heuristic solution to VR-CG resource allocation. However, this initial formulation addresses all components in a monolithic manner, which severely impacts scalability and hinders its application in large-scale or real-time scenarios, characteristics that are essential for 6G-enabled VR-CG services. Additionally, our previous work do not consider communication resource scheduling and the impact of network resource allocation in the perceived Motion-to-Photon latency.

Collectively, the reviewed studies represent meaningful progress toward enabling immersive applications over modern networks. Nonetheless, most are limited to local environments, rely on small scale networks, or lack adherence to 3GPP specifications relevant for emerging future network systems \cite{TR38838:2022}. Moreover, they do not address the scalability challenges critical for supporting dense user scenarios. To overcome these limitations, this work proposes a novel multi-stage resource allocation formulation for VR-CG, designed to operate over a CNC-based 6G network architecture. The formulation incorporates attention-aware semantic communication models to efficiently deliver rendered video frames from cloud servers to user HMDs. A summary of the related literature and a comparative analysis of their contributions are presented in Table~\ref{tab:related_work}, highlighting the key differentiating aspects of our proposal.

\begin{table*}
\centering
\footnotesize
\begin{tabular}{|c|c|c|c|c|c|c|c|}
\hline Reference            & Application    & User-centric QoE        & SC transmission         & 6G networks             & 3GPP compliant          & PRB scheduling                & Scalability              \\ \hline
\cite{Du_2023}       & Metaverse      & $\textcolor{blue}{\checkmark}$          & $\textcolor{blue}{\checkmark}$          & $\textcolor{red}{\times}$              & $\textcolor{red}{\times}$              & $\textcolor{red}{\times}$              & $30$ users             \\
\cite{Li:2020}       & VR-CG          & $\textcolor{red}{\times}$              & $\textcolor{red}{\times}$              & $\textcolor{red}{\times}$              & $\textcolor{red}{\times}$              & $\textcolor{red}{\times}$              & $10$~users             \\
\cite{Wang:2018}     & Social VR      & $\textcolor{red}{\times}$              & $\textcolor{red}{\times}$              & $\textcolor{red}{\times}$              & $\textcolor{red}{\times}$              & $\textcolor{red}{\times}$          & $300$~users            \\
\cite{Zou:2021}      & VR-CG          & $\textcolor{blue}{\checkmark}$          & $\textcolor{blue}{\checkmark}$          & $\textcolor{red}{\times}$              & $\textcolor{red}{\times}$              & $\textcolor{red}{\times}$              & $15$~users             \\
\cite{Lee:2024}      & VR-CG          & $\textcolor{blue}{\checkmark}$          & $\textcolor{red}{\times}$              & $\textcolor{red}{\times}$              & $\textcolor{red}{\times}$              & $\textcolor{red}{\times}$              & $20$~users             \\
\cite{Song:2023}     & VR-CG          & $\textcolor{blue}{\checkmark}$          & $\textcolor{blue}{\checkmark}$          & $\textcolor{red}{\times}$              & $\textcolor{red}{\times}$              & $\textcolor{red}{\times}$              & $17$~users             \\
\cite{Vaidya:2023}   & VR-CG          & $\textcolor{blue}{\checkmark}$          & $\textcolor{red}{\times}$              & $\textcolor{red}{\times}$              & $\textcolor{red}{\times}$              & $\textcolor{red}{\times}$              & Single user              \\
\cite{Zhao:2021}     & VR-CG          & $\textcolor{blue}{\checkmark}$          & $\textcolor{red}{\times}$              & $\textcolor{red}{\times}$              & $\textcolor{red}{\times}$              & $\textcolor{red}{\times}$              & Single user              \\
\cite{Warsinke:2024} & VR-CG          & $\textcolor{blue}{\checkmark}$          & $\textcolor{red}{\times}$              & $\textcolor{red}{\times}$              & $\textcolor{red}{\times}$              & $\textcolor{red}{\times}$              & Single user              \\
\cite{Zhang:2017}    & VR-CG          & $\textcolor{blue}{\checkmark}$          & $\textcolor{blue}{\checkmark}$          & $\textcolor{red}{\times}$              & $\textcolor{red}{\times}$              & $\textcolor{red}{\times}$          & $24$~users             \\
\cite{Almeida:2024}  & VR-CG          & $\textcolor{blue}{\checkmark}$          & $\textcolor{blue}{\checkmark}$          & $\textcolor{blue}{\checkmark}$          & $\textcolor{blue}{\checkmark}$          & $\textcolor{red}{\times}$          & $16$ users             \\ \hline
\textbf{This work}   & \textbf{VR-CG} & \textbf{$\textcolor{blue}{\checkmark}$} & \textbf{$\textcolor{blue}{\checkmark}$} & \textbf{$\textcolor{blue}{\checkmark}$} & \textbf{$\textcolor{blue}{\checkmark}$} & \textbf{$\textcolor{blue}{\checkmark}$} & \textbf{$\textbf{1700}$ users} \\
\hline
\end{tabular}
\caption{Related work comparison.}
\label{tab:related_work}
\end{table*}
\section{Virtual Reality Cloud Gaming System}
\label{sec:VR_CG_network}

In this section, we present the comprehensive VR-CG system considered in our work. We describe the VR-CG applications architecture, functionalities, and critical requirements, defining their operational flow and system demands. We then characterize the 6G network components and architectural elements involved in our resource allocation formulation. Finally, we introduce a novel attention-aware QoE model based on AI/ML models that integrate both visual quality and user interactivity in VR-CG.

\subsection{VR-CG Application Architecture}
\label{subsec:VRCG_arch}

\begin{figure}[t]
    \centering
    \includegraphics[width=\linewidth]{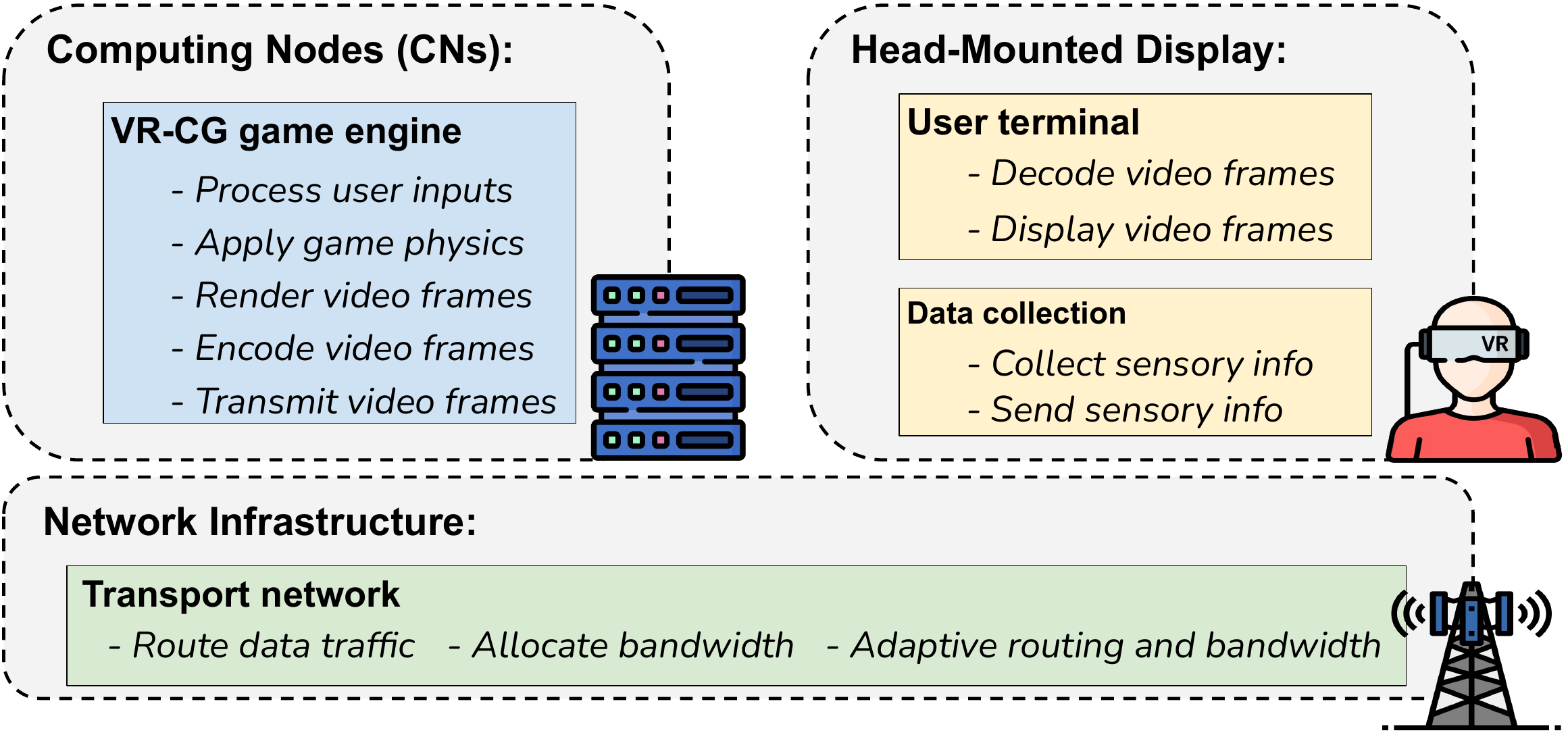}
    \caption{VR-CG application architecture.}
    \label{fig:VRCG_app_arch}
\end{figure}

The architecture of VR-CG builds upon the well established Cloud Gaming (CG) paradigm \cite{Shea:2013}. Similar to traditional CG systems, the VR-CG architecture consists of a cloud-based server infrastructure responsible for executing the game engine, which processes game logic and renders video frames to be transmitted to the user \cite{Alhilal:2024}. Complementing this, the user terminal is responsible for displaying the rendered frames and transmitting user input commands back to the cloud. The VR-CG application architecture is illustrated in Figure \ref{fig:VRCG_app_arch}.

A key benefit of CG, inherited by VR-CG, is enabling advanced gaming experiences on lightweight, low-power devices. However, VR-CG extends this paradigm by introducing immersive VR requirements. Similar to conventional gaming, VR applications also demand extremely high computing power. However, this demand is further challenged by the need to render ultra-high resolution images independently for both eyes to enable stereoscopic vision \cite{Ahmed:2017}. Achieving such performance locally on HMDs is infeasible due to limitations in size, weight, power consumption, and heat dissipation. VR-CG addresses these constraints by offloading computationally intensive tasks to remote servers within the network.

The VR-CG system comprises two disaggregated components: the VR-CG game engine and the user terminal \cite{Alhilal:2024}. The game engine runs on remote Computing Nodes (CNs), handling the virtual environment simulation of the virtual environment, game physics, user input processing, and ultra-high resolution rendering. The user terminal decodes and displays received video frames, while transmitting controller inputs, head movements, and HMD telemetry data back to the game engine \cite{Almeida:2025}.

The network plays a critical role in enabling this architecture, connecting the VR-CG game engine to the user’s HMD while meeting stringent requirements. The downlink must support high throughput transmission of ultra-high resolution frames, whereas the uplink must carry latency-sensitive input data to ensure synchronization between user actions and the virtual scene. As such, 3GPP classifies VR-CG traffic as Ultra-Reliable Low-Latency Communication (URLLC) on the uplink and Enhanced Mobile Broadband (eMBB) on the downlink \cite{TR22842:2019}. This dual requirement positions VR-CG as a particularly demanding use case for 6G networks.

\subsection{Cellular Network Environment}
\label{subsec:wireless_network}

To enable dynamic resource allocation across computing and communication domains, and to support adaptive placement of VR-CG game engines, the CNC architecture has emerged as a foundational element of 6G networks \cite{Tang:2024}. Conceptually similar to Multi-access Edge Computing (MEC) \cite{Fraga:2022}, CNC advances this paradigm by tightly integrating distributed computing resources with network control mechanisms, enabling holistic orchestration of services with stringent requirements.

\begin{figure}[t]
    \centering
    \includegraphics[width=0.98\linewidth]{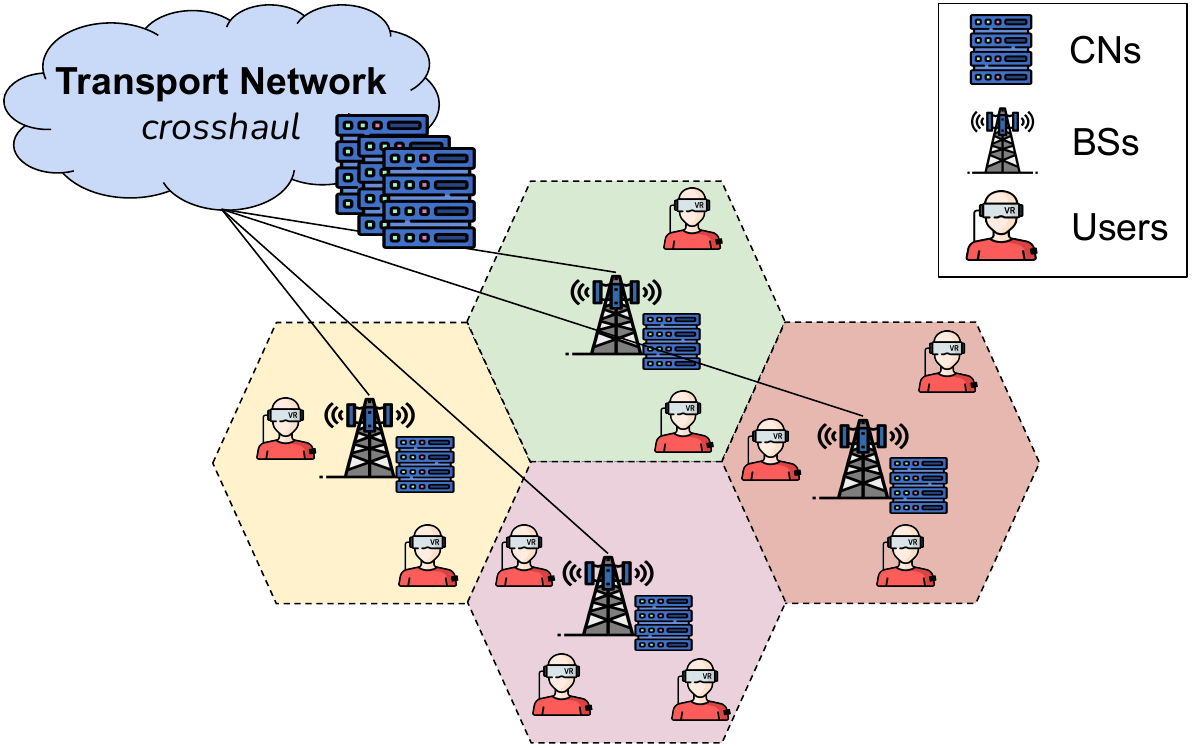}
    \caption{NG-RAN topology.}
    \label{fig:NG-RAN}
\end{figure}

The CNC architecture deploys a distributed set of CNs interconnected by a high-capacity, low-latency Transport Network (TN) \cite{Morais:2020}, as depicted in Figure~\ref{fig:NG-RAN}. CNs are strategically placed across the network, enabling trade-offs between computing proximity and resource availability. CNs may be co-located with Base Stations (BSs) for minimal latency or hosted in centralized datacenters for greater scalability. This allows the CNC to dynamically select the optimal placement of VR-CG game engines based on different criteria such as network conditions, user mobility, and resource availability.

In this work, we consider a Next Generation Radio Access Network (NG-RAN) architecture \cite{Morais:2020}, in accordance with 3GPP specifications \cite{TR38913:2024}. The NG-RAN operates in Frequency Range 1 (FR1) and employs Single User (SU) MIMO systems \cite{TR38838:2022}, with multiple antennas at both the transmitter and receiver to enhance wireless transmission. The BSs are interconnected via the TN, which also provides connectivity to the CNC’s distributed CNs. This architecture enables seamless communication between users' HMDs and the remotely deployed VR-CG game engines.

Wireless channel conditions have a significant impact on VR-CG users. Small-scale fading and co-channel interference can degrade link quality, increasing resource consumption to maintain target QoE \cite{Du_2023}. To address this, we explicitly incorporate channel state information in our resource allocation strategy, dynamically adapting bandwidth based on the channel quality between users and BSs. To represent the channel quality we calculate the Signal-to-Interference-plus-Noise Ratio (SINR) for user $u$ and stream $s$ as follows \cite{Tse:2005}:
\begin{equation}
\label{eq:sinr}
SINR(u, s) = \frac{\mathbf{h}_s^H \mathbf{p}_s \mathbf{h}_s}{\sum_{t \neq s} \mathbf{h}_t^H \mathbf{p}_t \mathbf{h}_t + (N_0 B)},
\end{equation}
\noindent where $\mathbf{h}_s$ is the channel vector for stream $s$, $\mathbf{p}_s$ the transmit power matrix, $B$ the allocated bandwidth, $N_0$ the thermal noise density, and $\mathbf{h}_s^H$ the Hermitian transpose of $\mathbf{h}_s$.

To further support VR-CG performance, we assume an NG-RAN architecture with Multi-Connectivity (MC) \cite{TS37340:2024,Cai:2021,Suer:2020}, enabling users to connect simultaneously to multiple BSs and receive parallel data streams. This enhances throughput and reliability, improving robustness against network fluctuations. To model wireless propagation in the VR-CG environment, we consider a dense urban system by adopting the Urban Microcell Street Canyon model \cite{TR38901:2024}. We define the Line-of-Sight ($P_l$) and Non-Line-of-Sight ($P_n$) path loss model as follows:
{
\small
\begin{equation}
    P_l(d) = 32.4 + 21 \log_{10}(d) + 20 \log_{10}(f)
\end{equation}
}%
{
\small
\begin{equation}
    P_n(d) = 35.3 + 22.4 \log_{10}(d) + 21.3 \log_{10}(f) - 0.3(h - 1.5),
\end{equation}
}%
\noindent where $d$ is the transmitter-receiver distance, $f$ is the carrier frequency, and $h$ is the user height.

\subsection{Attention-aware Resource Allocation}
\label{subsec:attention-aware}

Effective resource allocation for VR-CG in 6G networks requires a user-centric approach. QoE models must capture both objective visual quality and users' subjective interaction patterns. In VR-CG, users actively engage with virtual objects, manipulating them as part of gameplay. Consequently, resource allocation must dynamically adapt to user actions and visual attention to achieve high efficiency.
Given the high interactivity of VR-CG applications, users may exhibit varying levels of attention toward different virtual objects within their FoV. To capture this behavior, VR-CG service providers can store historical attention data linked to virtual objects across game sessions. However, a user may be engaging with the game for the first time or may not have previously encountered certain virtual objects, leading to sparsity in per-user attention data. To address this, it is possible to leverage aggregated historical data from other users, combined with semantic object segmentation, to predict a user's attention level for all virtual objects \cite{Du:2023}. This enables attention-aware resource allocation in the 6G network infrastructure.

\begin{figure}[t]
    \centering
    \includegraphics[width=\linewidth]{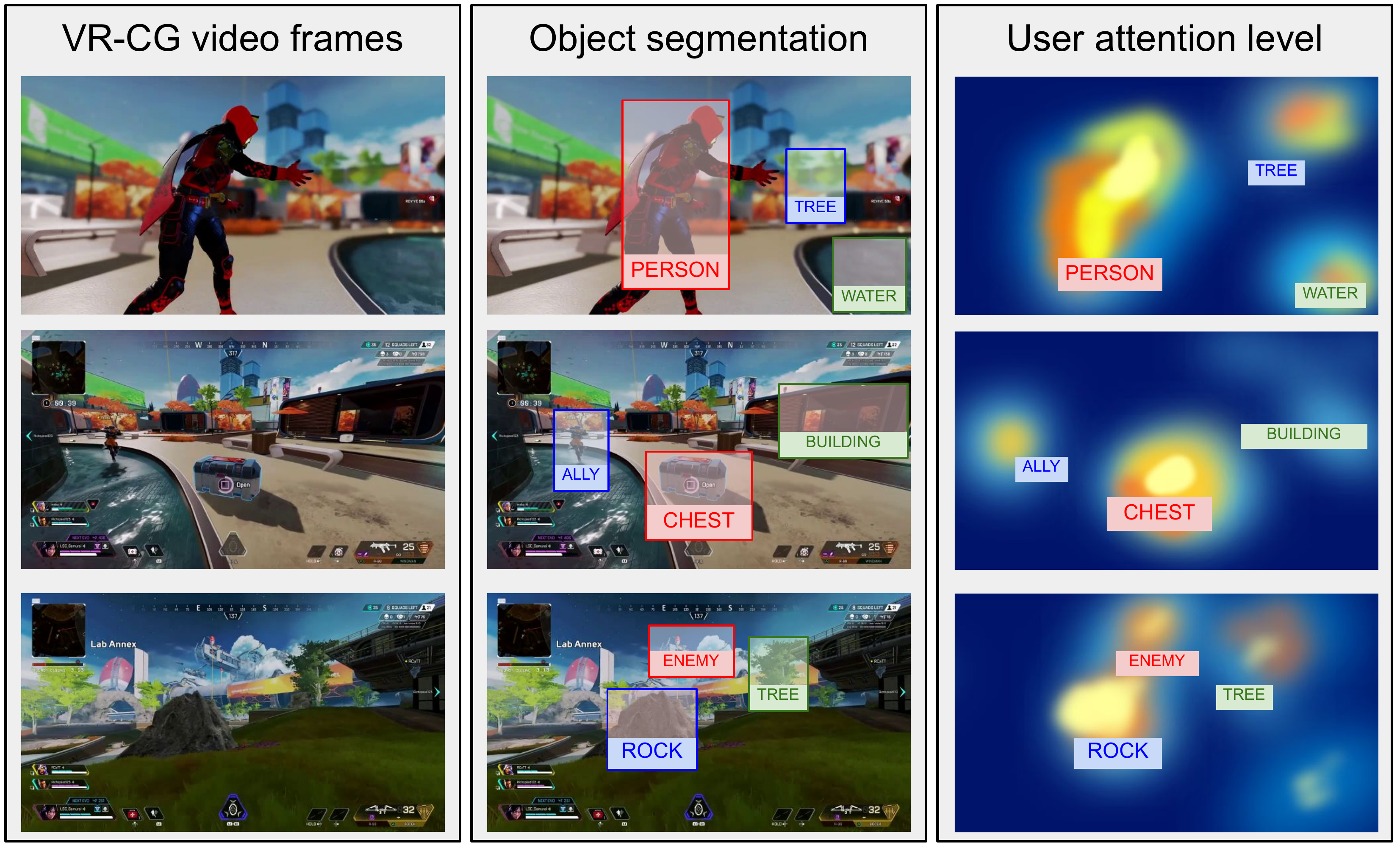}
    \caption{Attention-aware mechanism.}
    \label{fig:attention}
\end{figure}

To build our attention level mechanism, we leverage eye-tracking technologies and CV–based solutions to identify objects within the user's FoV in the virtual environment. For object segmentation, we adopt the approach proposed in \cite{Du_2023}, employing K-Net \cite{Zhang:2021}, a state-of-the-art semantic segmentation algorithm, to assign object labels and determine pixel-level positions of each virtual object. After segmentation, we apply the AI/ML-based attention prediction mechanism introduced in \cite{Du_2023} to estimate the user's attention level toward each object. This prediction is informed by the user profile dataset described in \cite{Du:2023}. The object segmentation mechanism and user attention level data are illustrated in Figure \ref{fig:attention}.

VR-CG naturally aligns with the principle of SC, which enable network systems to prioritize the transmission of semantically relevant content. This paradigm is particularly advantageous in VR-CG scenarios, where user attention shifts dynamically across the virtual environment. By focusing on transmitting only the most contextually significant information, SC helps mitigate the stringent latency and bandwidth requirements inherent to immersive VR-CG experiences. In this work, we adopt the object-aware video transmission model \cite{Du:2023}, tailoring it to the VR-CG context. The model leverages semantic object segmentation and user attention predictions to compute a semantic feature, which guides transmission resource allocation. In this transmission framework, video frames are decomposed into object-centric image segments, with each object transmitted independently and encoded at a resolution proportional to its estimated attention level \cite{Zou:2021,Song:2023}.

This approach departs from conventional video transmission that uniformly encodes and transmits entire ultra-high resolution frames. Instead, the proposed framework selectively transmits only the image fragments corresponding to the pixels of individual virtual objects. This strategy enables substantial savings in communication resources compared to the baseline of transmitting full frames at uniformly high resolution \cite{Du_2023}.

\begin{figure}[t]
    \centering
    \includegraphics[width=\linewidth]{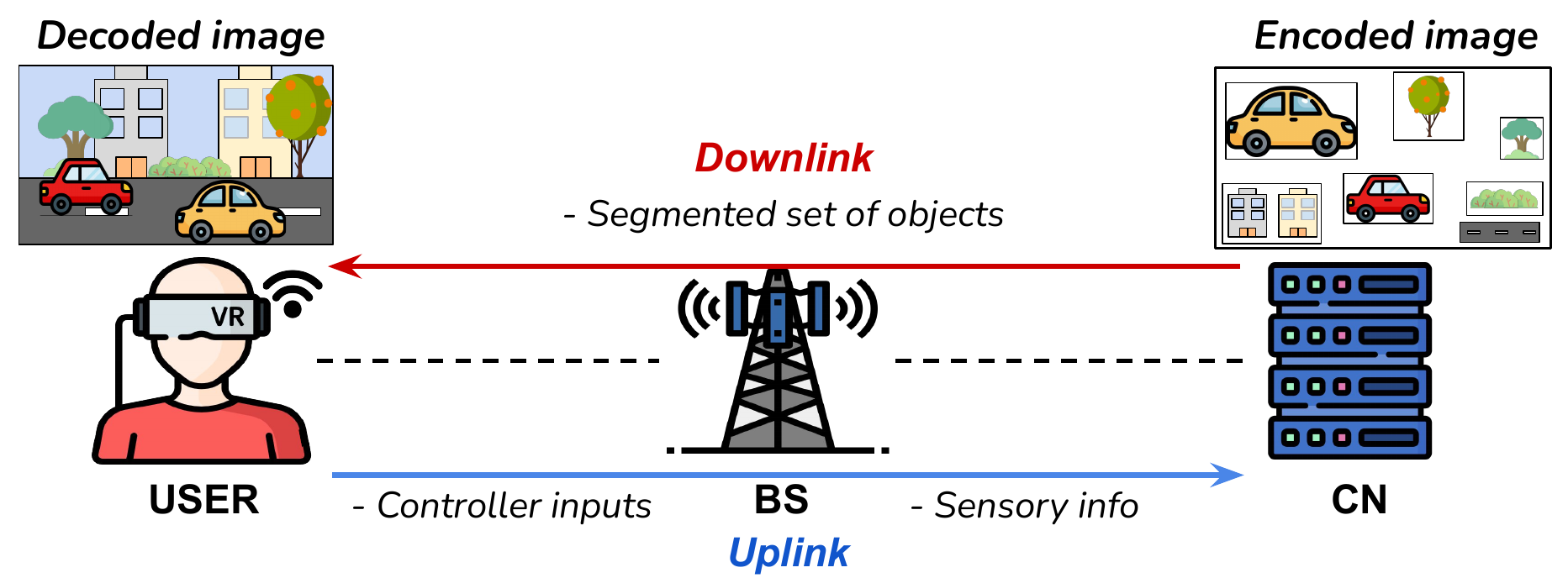}
    \caption{VR-CG object-aware transmission model.}
    \label{fig:attention_transmission_model}
\end{figure}

Figure \ref{fig:attention_transmission_model} illustrates the proposed object-aware transmission model for VR-CG applications. The CNs render and encode virtual objects independently, considering the predicted user attention levels. Consequently, the CN transmits a set of virtual object image segments over the 6G network. At the user side, the HMD decodes the received segments and reconstructs the full virtual scene by compositing and upscaling the object images to match the HMD's display resolution. Finally, the fully composed and decoded virtual scene is presented to the user.

Following the defined VR-CG transmission system, we propose a QoE model for VR-CG applications that accounts for both objective image quality metrics and users’ subjective preferences. The model incorporates traditional image quality indicators to capture user preferences tied to the specific game type and its visual demands. Additionally, it integrates attention-aware metrics that reflect users’ perceptual focus and interaction patterns within the virtual environment.

To guide efficient resource allocation while ensuring fairness among users, we base our QoE formulation on the Weber–Fechner Law \cite{Dehaene:2003}. This introduces a logarithmic relationship between perceived satisfaction and allocated resources, effectively modeling the diminishing returns of QoE improvements as quality increases. The formulation inherently promotes fairness by prioritizing enhancements for users experiencing lower QoE, rather than allocating disproportionate resources to those already enjoying high quality. This balance ensures efficient resource allocation while preserving a user-centric perspective. We calculate the QoE of user $u$ as follows:
\begin{equation}
\label{eq:QoE_model}
    \mathcal{Q}(u) = \sum_{o \in \mathcal{O}} \left [ \lambda(u, o) \ln \left ( \frac{\rho(o) \tau(f)}{\min\limits_{\rho, \tau}(\rho(o) \tau(f))} \right ) \right ],
 \end{equation}
\noindent where $\mathcal{O}$ denotes the set of virtual objects, $\lambda_u^o$ is the attention level of user $u$ to object $o$, $\rho(o)$ is the resolution quality coefficient of object $o$, and $\tau(f)$ is the frame rate quality coefficient based on the selected refresh rate $f$.

Building on this semantic and attention-aware transmission paradigm, the following section formalizes the system model underlying our VR-CG framework. We detail the architecture components, network assumptions, and resource constraints considered throughout this work.
\section{System Model and Problem Formulation}
\label{sec:system_model}

In this section, we formalize the network and application environment that supports the execution and delivery of VR-CG services, including the underlying infrastructure and user interaction parameters. We assume a 6G network infrastructure that provides wireless cellular connectivity to support user access to VR-CG applications. The network is integrated with a CNC architecture that hosts the VR-CG game engines and associated services.

Following the 3GPP specifications \cite{TS38300:2024, TS38401:2024}, we consider a NG-RAN system integrated with a 6G core network. The network topology is illustrated in Figure \ref{fig:NG-RAN} and consists of the following primary components:

\begin{itemize}
    \item A set $\mathcal{B}$ with all BSs in the RAN topology, where each element $b \in \mathcal{B}$ denotes a BS serving as an access point for wireless connectivity. Each BS has a set of Physical Resource Blocks (PRBs) representing its bandwidth resource pool, denoted as $b^{PRBs}$.
    \item A set $\mathcal{U}$ of users, where each user $u \in \mathcal{U}$ is equipped with an HMD to access VR-CG applications.
    \item A set $\mathcal{C}$ of CNs empowered by a CNC architecture. Each CN $c \in \mathcal{C}$ is characterized by multiple computing resources, including processing $c^{CPU}$, rendering $c^{GPU}$, memory $c^{RAM}$, and network $c^{NET}$ capacity.
    \item A set $\mathcal{P}_b^c$ of paths representing the available paths connecting each BS $b \in \mathcal{B}$ to each CN $c \in \mathcal{C}$. Each path $p \in \mathcal{P}_b^c$ consists of a sequence of directed links $e_{i, j}$, where each link is characterized by a bandwidth capacity denoted as $e_{i, j}^{Cap}$. The transmission latency of a path $p \in \mathcal{P}_b^c$, denoted by $l(p)$, is computed as the sum of the latencies of all constituent links.
\end{itemize}

According to the VR-CG system described in Section~\ref{sec:VR_CG_network}, we define the following VR-CG environment elements:

\begin{itemize}
    \item A set $\mathcal{G}$ with all available VR-CG video games that users can access in the system. Each video game $g \in \mathcal{G}$ corresponds to a specific VR-CG application, with its demand configuration and game type.
    \item A set $\mathcal{R}_u$ of supported display resolutions for each user $u \in \mathcal{U}$. Each element $r \in \mathcal{R}_u$ represents a resolution option that can be selected for the user.
    \item A set $\mathcal{F}_u$ of supported frame rates options for  each user $u \in \mathcal{U}$. Each $f \in \mathcal{F}_u$ represents an available refresh rate option for the user.
\end{itemize}
VR-CG resource allocation represents a joint optimization problem that simultaneously addresses multiple interdependent dimensions of wireless communication and cloud-based services. It includes user association, wireless resource allocation, VR-CG game engine placement, transport network routing, and video quality adaptation. Each of these components must be addressed under stringent performance and resource constraints.

The user association component aims to assign each user to an appropriate BS based on wireless channel quality and resource availability. The goal is to ensure reliable wireless connectivity while distributing the communication load efficiently. The application placement task involves selecting a suitable CN to host the VR-CG game engine for each user and determining feasible transport paths for delivering VR frames. This requires balancing computing loads across CNs and ensuring that the selected network paths respect bandwidth constraints. Finally, the video quality adaptive selection focuses on choosing image quality metrics that enhance user QoE without exceeding the limits of the communication and processing infrastructure.

Each of these subproblems is NP-hard and interrelated \cite{Morais:2020,Liu:2016,Campos:2024}. Attempting to solve them jointly in a monolithic optimization model leads to intractable complexity, especially under dynamic conditions such as user mobility and time-varying network states \cite{Almeida:2025}.
To tackle this high complexity, we adopt a hierarchical multi-stage optimization framework in which each stage focuses on a distinct subproblem of VR-CG resource allocation. The first stage addresses user association and the coarse-grained allocation of wireless communication resources, ensuring efficient utilization of the NG-RAN resources. The second stage focuses on the dynamic placement of VR-CG game engines, allocation of computing resources, and selection of transport paths across the network, considering link and processing constraints. The third stage introduces a user-centric, attention-aware model that dynamically adjusts video resolution and frame rate based on user attention level to virtual objects and PRB scheduling, i.e., the fine-grained wireless resource allocation.

\subsection{First stage problem formulation}
\label{subsec:first_stage}

In the first stage, we focus on the subproblem of user association and wireless transmission resource allocation, which forms the foundation of the overall VR-CG resource management. This stage determines the optimal assignment of users to base stations and allocates bandwidth resources to satisfy their communication requirements. Additionally, an initial selection of image resolution and frame rate is performed at this stage. This initial choice serves as a baseline and does not yet incorporate user attention awareness within the virtual environment, which will be addressed later in the third stage.

The decision variables considered in the first stage are:

\begin{itemize}
    \item $x_{u}^{b} = \{0, 1\}$: A binary variable that indicates if the user $u \in \mathcal{U}$ is associated with BS $b \in \mathcal{B}$.
    \item $y_{u}^{b} \in \mathbb{Z}_{\geq 0}$: An integer variable that represents the amount of PRBs allocated in BS $b \in \mathcal{B}$ to user $u \in \mathcal{U}$.
    \item $w_u^r = \{0, 1\}$: A binary variable that indicates if resolution $r \in \mathcal{R}_u$ is selected to user $u \in \mathcal{U}$.
    \item $z_u^f = \{0, 1\}$: A binary variable that indicates if frame rate $f \in \mathcal{F}_u$ is selected to user $u \in \mathcal{U}$.
    \item $o_u^b \in \mathbb{R}$: A continuous variable that indicates the proportion of data that is transmitted from BS $b \in \mathcal{B}$ to user $u \in \mathcal{U}$.
\end{itemize}

The first stage goal is to determine optimal user associations and to allocate wireless transmission resources efficiently, ensuring that the bandwidth and latency requirements of VR-CG game engines are met. This stage also considers users' gaming preferences, which play a crucial role in their perceived QoE. For instance, modern gaming platforms such as the PlayStation 5 \cite{Bangash:2024} offer two standard experience modes: (i) resolution mode, which favors high graphical fidelity at the expense of lower frame rates, and (ii) performance mode, which prioritizes higher frame rates with reduced visual quality.

Motivated by this distinction, we assume that each VR-CG game $g \in \mathcal{G}$ is associated with a default gaming mode that reflects its intended user preference. Specifically, we define two preference modes: the quality mode, which prioritizes higher resolution and visual fidelity, and the performance mode, which emphasizes smoother gameplay through higher frame rates. We introduce an input parameter $P(u) = \{0, 1\}$ indicating if user $u \in \mathcal{U}$ is playing a quality mode game ($P(u) = 1$) or a performance mode game ($P(u) = 0$). This preference can be derived from the VR-CG video game settings, as discussed in Section \ref{subsec:experimental_setup}.

The objective of the first stage is to maximize the overall QoE across all users. We formulate the objective function as the sum of individual user QoE values, capturing the impact of user association, resource allocation, and video quality selection. The QoE of each user is calculated following the proposed VR-CG QoE model, defined in Equation (\ref{eq:QoE_model}), in conjunction with the game-specific preference parameter $P(u)$, but without the user attention awareness ($\lambda_u^o$) that will be addressed later in the third stage. Thus, the objective function is defined as:
\begin{align}
    \maximize_{w_u^r, z_u^f} \sum_{u \in \mathcal{U}} \Bigg[ & P(u) \sum_{r \in \mathcal{R}_u} \ln \left( \frac{w_u^r}{\min\limits_{r}(\mathcal{R}_u)} \right) + \nonumber \\
    & (1 - P(u)) \sum_{f \in \mathcal{F}_u} \ln \left( \frac{z_u^f}{\min\limits_{f}(\mathcal{F}_u)} \right) \Bigg].
\end{align}

To meet the bandwidth and latency demands of VR-CG applications, users are allowed to connect to multiple BSs simultaneously, improving robustness and load balancing in Multi-Connectivity systems \cite{TS37340:2024}. However, due to limitations in wireless communication systems, a maximum of $N$ simultaneous connections per user is imposed. The following constraint ensure that each user is associated with at least one, but no more than $N$ BSs:
\begin{equation}
    1 \leq \sum_{b \in \mathcal{B}} x_u^b \leq N, \qquad \forall u \in \mathcal{U}.
\end{equation}

Following user association, PRBs are allocated to enable wireless communication between users' HMDs and VR-CG game engines. The following constraints ensure that each user receives a positive amount of PRBs exclusively from the BSs to which they are associated:
\begin{equation}
    y_u^b \geq x_u^b, \qquad \forall u \in \mathcal{U}, b \in \mathcal{B}.
\end{equation}
\begin{equation}
    x_u^b b^{PRBs} \geq y_u^b, \qquad \forall u \in \mathcal{U}.
\end{equation}

Additionally, it is essential to ensure that the communication resource capacity of each BS is not exceeded:
\begin{equation}
    \sum_{u \in \mathcal{U}} y_u^b \leq b^{PRBs}, \qquad \forall b \in \mathcal{B}.
\end{equation}

To guarantee a seamless VR-CG experience, the following constraints defines that each user is assigned to exactly one video resolution and one frame rate:
\begin{equation}
    \sum_{r \in \mathcal{R}} w_u^r = 1, \qquad \forall u \in \mathcal{U}.
\end{equation}
\begin{equation}
    \sum_{f \in \mathcal{F}} z_u^f = 1, \qquad \forall u \in \mathcal{U}.
\end{equation}

Given the selected resolution and frame rate, the VR-CG transmission load can be estimated based on the number of pixels, compression efficiency, and frame rate. We calculate the total traffic load for the user $u \in \mathcal{U}$ to BS $b \in \mathcal{B}$ as follows:
\begin{equation} 
    Load(u, b) = o_u^b \sum_{r \in \mathcal{R}_u} \left( \Phi(w_u^r) \cdot \eta \right) \cdot \sum_{f \in \mathcal{F}} \Psi(z_u^f),
\end{equation}
\noindent where $\Phi(w_u^r)$ denotes the number of pixels of the selected image resolution, $\eta$ is the compression rate applied by the encoding scheme, and $\Psi(z_u^f)$ indicates the selected number of frames per second (FPS).

We estimate the achievable throughput based on PRBs allocation to each user. Let $\omega(b)$ be the PRBs bandwidth size of BS $b \in \mathcal{B}$. Using Shannon's capacity equation \cite{Almeida:2025}, we calculate the maximum achievable data rate for user $u \in \mathcal{U}$ and BS $b \in \mathcal{B}$, as follows:
\begin{equation}
\label{eq:shannon}
    TP(u, b) = y_u^b \omega(b) \log_2 \Bigg( 1 + SINR(u, b) \Bigg).
\end{equation}

In this way, we can formulate the VR-CG throughput constraint as follows:
\begin{equation} 
\ Load(u, b) \leq TP(u, b), \qquad \forall u \in \mathcal{U}, b \in \mathcal{B}. 
\end{equation}

The total end-to-end latency for transmitting video frames from the VR-CG game engine to the user's HMD is modeled by considering six key components: (i) routing latency, which accounts for the time to forward packets from the CN to the serving BS; (ii) rendering latency, representing the processing delay to generate and encode video frames at the CN; (iii) propagation latency, reflecting the physical travel time from the BS to the user's HMD; (iv) transmission latency over the wireless channel; (v) BS processing latency for decoding, scheduling, and forwarding packets; and (vi) buffer latency caused by congestion or scheduling delays at the BS.

Since the first stage does not explicitly optimize VR-CG game engine placement, we assume each BS uses its nearest CN, denoted by $\mathbf{c}_b$, to host the game engine, following \cite{Esper:2023}. Consequently, all users associated with a BS are served by the game engine at this CN, and the first-stage routing latency is expressed as:
\begin{equation}
\label{eq:routing_latency}
\mathbf{R}_{1st}(u, \mathbf{c}_b) = \max \Big \{ x_u^b \ \mathbf{l}(p) : b \in \mathcal{B}, p \in \mathcal{P}_b^{\mathbf{c}_b} \Big \},
\end{equation}
where $\mathbf{l}(p)$ represents the latency of the $p \in \mathcal{P}$.

Rendering latency is the time taken by the VR-CG application to generate a fully rendered frame before transmission, involving tasks such as scene composition, texture mapping, lighting, and encoding, all of which become more time-consuming at higher resolutions and frame rates. To model this, let $\mathbf{r}(c)$ denote the rendering speed (in pixels per second) of CN $c \in \mathcal{C}$. Given the selected resolution $\Phi(w_u^r)$ and frame rate $\Psi(z_u^f)$, the rendering latency for the user $u \in \mathcal{U}$, when served by CN $\mathbf{c}_b$, is calculated as:
\begin{equation}
\label{eq:render_latency}
\mathbf{S}(u, b) = \sum_{f \in \mathcal{F}} \sum_{r \in \mathcal{R}} \Bigg ( \frac{x_u^b \Phi(w_u^r) \Psi(z_u^f)}{\mathbf{r}(\mathbf{c}_b)} \Bigg ).
\end{equation}

Propagation latency is the time it takes for electromagnetic signals to travel from the BS to the user's HMD, which depends on the distance $\mathbf{d}(u, b)$ between them and the speed of light $L$. This latency is modeled as:
\begin{equation}
\label{eq:prop_latency}
\mathbf{P}(u, b) = \frac{x_u^b \mathbf{d}(u, b)}{L}.
\end{equation}

Transmission latency is the time needed to send a VR-CG frame over the wireless channel, and is determined by the frame size $\mathbf{s}(\Phi(w_u^r))$ in bits and the user's available throughput $TP(u, b)$. It is given by:
\begin{equation}
\label{eq:trans_latency}
\mathbf{T}(u, b) = \sum_{r \in \mathcal{R}} \Bigg (\frac{x_u^b \mathbf{s}(\Phi(w_u^r))}{TP(u, b)} \Bigg ).
\end{equation}

The BS's processing latency accounts for the time to decode, schedule, and schedule resources for the VR-CG frame packets before forwarding. This latency is influenced by the size of the video frame and the computing capacity $\rho(b)$ of BS $b \in \mathcal{B}$. We formulate it as:
\begin{equation}
\label{eq:proc_latency}
\mathbf{O}(u, b) = \sum_{r \in \mathcal{R}} \left ( \frac{x_u^b \mathbf{s}(\Phi(w_u^r))}{\rho(b)} \right ).
\end{equation}

Buffer latency arises when incoming packets wait in line at the BS due to constraints in processing or transmission capacity. Although it varies dynamically with network traffic, we estimate it using an analytical approximation based on the user's packet arrival rate $\mathbf{a}_u$ and the BS's processing capacity $\rho(b)$. It is expressed as:
\begin{equation}
\label{eq:queue_latency}
\mathbf{Q}(u, b) = x_u^b \left ( \rho(b) - \sum\limits_{u' \in \mathcal{U}} \mathbf{a}_{u'} \right )^{-1}.
\end{equation}

To ensure a high-quality and immersive VR-CG experience, each video frame must be delivered to the user's HMD within a strict latency bound. VR-CG applications depend on low latency rendering and transmission. Therefore, any excessive delay can severely impact user experience by inducing motion sickness, reducing responsiveness, and breaking the sense of presence \cite{Xia:2023,TR38838:2022}.

We define the end-to-end (e2e) latency experienced by each user as the sum of all latency components involved in the communication process. We calculate e2e latency for user $u \in \mathcal{U}$ as follows: 
\begin{align}
    \mathbf{L}(u) =& \max \Big \{ \mathbf{R}_{1st}(u, \mathbf{c}_b) + \mathbf{S}(u, b) + \mathbf{P}(u, b) + \nonumber \\ & \mathbf{T}(u, b) + \mathbf{O}(u, b) + \mathbf{Q}(u, b) : b \in \mathcal{B} \Big \}.
\end{align}

The selected frame rate for a given user determines how many VR-CG video frames must be delivered to the user's HMD within a fixed time interval, typically 1 second. To maintain a smooth and responsive experience, every generated frame must be transmitted within this deadline, denoted by $\Delta^{max}$. Therefore, the latency constraint ensuring that the e2e delay for each user does not exceed this time window is defined as follows:
\begin{equation}
    \mathbf{L}(u) \leq \Delta^{max}.
\end{equation}

By solving the first stage subproblem, we establish an initial configuration that provides a feasible and structured foundation for the subsequent stages of the optimization framework. This stage ensures valid user associations and preliminary resource allocation, enabling a consistent and efficient starting point for further refinement. In the next section, we present the second stage, which addresses the optimal placement of the VR-CG game engine and the dynamic routing of traffic flows within the transport network.

\subsection{Second stage problem formulation}
\label{subsec:second_stage}

In the second stage of our optimization framework, we address the subproblem of VR-CG application placement and routing within the transport network. This stage complements the first-stage solution by optimizing the allocation of computing resources in a dynamic placement scenario. Furthermore, it identifies optimal routing paths across the transport network, leveraging multi-path routing to enhance both performance and resilience. The formulation of the second-stage problem introduces the following decision variables:
\begin{itemize} 
    \item $v_u^c = \{0, 1\}$: A binary variable indicating if CN $c \in \mathcal{C}$ hosts the VR-CG game engine for user $u \in \mathcal{U}$. 
    \item $s_u^p = \{0, 1\}$: A binary variable indicating if path $p \in \mathcal{P}_b^c$ is selected to route the traffic to user $u \in \mathcal{U}$.
    \item $q_u^p \in [0, 1]$: A continuous variable representing the proportion of traffic flow to user $u \in \mathcal{U}$ that is transmitted through path $p \in \mathcal{P}_b^c$.
\end{itemize}

The objective of the second-stage problem formulation is to optimize cost efficiency while preserving the QoE levels achieved in the first stage. Following the methodology proposed in \cite{Almeida:2024}, we define the cost of placing VR-CG game engines by considering three categories: (i) a fixed cost, incurred for keeping a CN powered on regardless of workload; (ii) a variable cost, which reflects the operational expenses associated with the actual consumption of computing resources; and (iii) a migration cost, computed based on the relocation of game engines from their previous placement to a new location.

We denote $\mathbf{v}^F_c$ as the fixed cost of activating CN $c \in \mathcal{C}$. The fixed cost considers whether the CN is activated or not, independently of its resource usage. In this way, we denote as $\gamma^F_c$ the cost to activate the CN $c \in \mathcal{C}$. The fixed cost of a CN  $c \in \mathcal{C}$ is calculated as follows:
\begin{equation}
    \mathbf{v}^F_c = 
    \begin{cases}
        \gamma^F_c,& \text{if} \sum\limits_{u \in \mathcal{U}} v_u^c \geq 1.\\
        0,& \text{otherwise}.
    \end{cases}
\end{equation}

To compute the variable cost, we first determine the resource utilization of each CN based on its computing load. Accordingly, we denote $\mathbf{G}(u)$ as the rendering load generated by user $u \in \mathcal{U}$, $\mathbf{C}(u)$ as the processing load, $\mathbf{M}(u)$ as the memory load, and $\mathbf{N}(u)$ as the network interface load. Each of these parameters reflects the respective resource demand imposed by the user on the CN. The variable cost of a CN can then be represented as a function of its total computing load as follows:
{
\small
\begin{equation}
    \mathbf{v}^V_c = \sum\limits_{u \in \mathcal{U}} v_u^c \Bigg ( \mathbf{G}(u) \gamma^{{G}}_c  + \mathbf{C}(u)\gamma^{{C}}_c + \mathbf{M}(u)\gamma^{{M}}_c + \mathbf{N}(u)\gamma^{{N}}_c \Bigg ),
\end{equation}
}%
\noindent where $\gamma^{{G}}_c$, $\gamma^{{C}}_c$, $\gamma^{{M}}_c$, and $\gamma^{{N}}_c$ represent the cost per unit of GPU, CPU, memory, and network resources, for CN $c \in \mathcal{C}$.

We denote the migration cost of a user as $\mathbf{v}^M_u$, and represent the current placement of the game engine for user $u \in \mathcal{U}$ as $\mathbf{c}_u$. This allows the optimization model to detect whether the placement of the game engine has changed and to compute the corresponding migration cost. The migration cost incurred when relocating the game engine between CNs $\mathbf{c}_u, c \in \mathcal{C}$ is denoted by $\Omega_{\mathbf{c}_u}^{c}$. The migration cost for user $u \in \mathcal{U}$ is calculated as follows:
\begin{equation}
    \mathbf{v}^M_u = \sum_{c \in \mathcal{C}} \bigg ( v_u^c
        \Omega_{\mathbf{c}_u}^c \bigg ).
\end{equation}

We define the objective function of the second stage as the minimization of the total cost, expressed as follows:
\begin{equation}
    \minimize_{v_u^c} \sum_{c \in \mathcal{C}} \Bigg ( \mathbf{v}^F_c + \mathbf{v}^V_c \Bigg ) + \sum_{u \in \mathcal{U}} \mathbf{v}_u^M.
\end{equation}

The first constraint guarantees that each user is assigned to exactly one CN to host their VR-CG game engine instance and is defined as follows:
\begin{equation}
    \sum_{c \in \mathcal{C}} v_u^c = 1, \qquad \forall u \in \mathcal{U}.
\end{equation}

To avoid service degradation, we impose resource capacity constraints, represented as follows:
\begin{align}
    \sum_{u \in \mathcal{U}} v_u^c \mathbf{G}(u) \leq c^{GPU}, \qquad \forall c \in \mathcal{C},\\
    \sum_{u \in \mathcal{U}} v_u^c \mathbf{C}(u) \leq c^{CPU}, \qquad \forall c \in \mathcal{C},\\
    \sum_{u \in \mathcal{U}} v_u^c \mathbf{M}(u) \leq c^{RAM}, \qquad \forall c \in \mathcal{C},\\
    \sum_{u \in \mathcal{U}} v_u^c \mathbf{N}(u) \leq c^{NET}, \qquad \forall c \in \mathcal{C}.
\end{align}

In the second stage, we refine the VR-CG game engine placement by revisiting the initial solution from the first stage. We consider that all network traffic either originates from or is destined to the VR-CG game engine, focusing mainly on the downlink scenario, which is the most demanding. Rather than imposing a strict upper bound on routing latency, we compute it explicitly based on the selected routing paths, enabling a more accurate representation of the transport network. 

To address the stringent transmission requirements of VR-CG and reduce the impact of heavy traffic loads, we employ multi-path routing in the transport network \cite{Almeida:2022}. To model multi-path routing between users' HMDs and CNs, we denote $\mathbf{B}_u$ as the set of BSs serving user $u \in \mathcal{U}$. This information, derived from the first stage solution, serves as input for the second stage. To ensure that each traffic flow transmitted from the VR-CG game engine to each user HMD is routed through at least one path, we impose the following constraint:
\begin{equation}
    \sum_{p \in \mathcal{P}_{b}^c} s_u^p \geq v_u^c, \qquad \forall u \in \mathcal{U}, \ c \in \mathcal{C}, b \in \mathbf{B}_u.
\end{equation}

Furthermore, we impose consistency constraints to ensure that if a path $p \in \mathcal{P}_b^c$ is selected, then a positive portion of the traffic flow must be transmitted through this path. Conversely, if a path is not selected, it must not carry any traffic. To control traffic distribution granularity, we introduce a parameter $0 \leq \varepsilon < 1$, which provides a balanced and efficient load distribution. The following constraints formalize the consistency of path usage:
\begin{equation}
    s_u^p \geq q_u^p, \qquad \qquad \ \ \ \forall c \in \mathcal{C}, \ u \in \mathcal{U}, \ b \in \mathbf{B}_u, \ p \in \mathcal{P}_{b}^c,
\end{equation}
\begin{equation}
    s_u^p - q_u^p \leq 1 - \epsilon, \qquad \forall c \in \mathcal{C}, \ u \in \mathcal{U}, \ b \in \mathbf{B}_u, \ p \in \mathcal{P}_{b}^c.
\end{equation}

To ensure routing feasibility, we must guarantee that the total traffic load transmitted through each selected path does not exceed the available capacity of its links. We denote as $E$ the set of all links $e_{i, j}$ in the network topology. In this way, we define the capacity constraint of the links as follows:
{\small
\begin{equation}
    \sum_{u \in \mathcal{U}} \sum_{b \in \mathbf{B}_u} \sum_{c \in \mathcal{C}} \sum_{p \in \mathcal{P}_b^c} \Big ( v_u^c q_u^p \ TP(u, b) \Big ) \leq e_{i,j}^{Cap}, \forall e_{i, j} \in E.
\end{equation}
}%

To ensure that the entire traffic load generated by the VR-CG video frames is successfully transmitted to the user's HMD, we impose a traffic conservation constraint. This constraint is formulated as:
\begin{equation}
    \sum_{p \in \mathcal{P}_{b}^c} q_u^p = 1, \qquad \forall u \in \mathcal{U}, b \in \mathbf{B}_u.
\end{equation}

To determine the transport network routing latency in a multi-path routing scenario, we adopt a conservative approach by considering the latency of the selected path with the highest delay. This ensures that the worst-case latency is accounted for, preventing potential violations of latency constraints. We represent the routing latency of the user $u \in \mathcal{U}$, served by BS $b$, to the CN $c \in \mathcal{C}$ as follows:
\begin{equation}
    \mathbf{R}_{2nd}(u, c) = \max\{ s_u^p \ \mathbf{l}(p) : p \in \mathcal{P}_{b}^c, b \in \mathbf{B}_u\}.
\end{equation}

In the first stage, routing latency is estimated assuming the VR-CG game engine runs at the closest CN to each BS serving the users. However, always placing the game engine at the nearest CN can be cost-inefficient, as edge processing nodes closer to the BS tend to have higher operational costs. While such placements reduce routing latency, they increase overall network expenses. To balance this trade-off, the second stage refines the VR-CG game engine placement while ensuring that user latency constraints are still satisfied.

We denote $\mathbf{L}(u)$ as the e2e latency from the first stage solution and $\mathbf{R}(u, \mathbf{c}_b)$ as the routing latency calculated in that stage. To maintain latency within acceptable limits, we impose the following constraint:
\begin{equation}
    v_u^c \big (\mathbf{L}(u) - \mathbf{R}_{1st}(u, \mathbf{c}_b) + \mathbf{R}_{2nd}(u, c) \big ) \leq \Delta^{max}, \forall u \in \mathcal{U}, c \in \mathcal{C}.
\end{equation}

The second stage refines VR-CG game engine placement and optimizes traffic routing in the vRAN transport network, ensuring efficient resource use while maintaining the QoE set in the first stage. The next section presents the third stage, which focuses on adjusting frame rate and image resolution based on user attention. This user-centric approach enhances perceptual quality and resource efficiency, delivering a more immersive VR-CG experience.

\subsection{Third stage problem formulation}
\label{subsec:third_stage}

In the third and final stage, we focus on fine-grained wireless resource allocation using a user-centric QoE approach presented in Section \ref{sec:VR_CG_network}. This stage revisits key decisions from the first stage, considering the object-aware transmission model proposed in \cite{Du:2023}.

We use Transmission Time Interval (TTI) as the time unit for wireless resource allocation. We denote $\mathcal{K}$ as the set of TTIs, where each $k \in \mathcal{K}$ represents a TTI. Additionally, we define $\mathcal{O}_u$ as the set of virtual objects visible within the FoV of user $u \in \mathcal{U}$, where each $o \in \mathcal{O}_u$ represents an individual virtual object. We consider the following decision variables in the third stage formulation:
\begin{itemize}
    \item $w_{u}^{r, o} = \{0, 1\}$: A binary variable that indicates if the resolution $r \in \mathcal{R}_u$ is selected to object $o \in \mathcal{O}_u$ of user $u \in \mathcal{U}$.
    \item $r_{u}^{b, k} = \{0, 1\}$: A binary variable that indicates if the BS $b \in \mathbf{B}_u$ transmits data to the user $u \in \mathcal{U}$ during TTI $k \in \mathcal{K}$.
    \item $n_{u}^{b, k} \in \mathbb{Z}_{\geq 0}$: An integer variable that indicates the amount of PRBs allocated to user $u \in \mathcal{U}$ during TTI $k \in \mathcal{K}$ in BS $b \in \mathbf{B}_u$.
\end{itemize}

The goal of the third-stage formulation is to maximize users' QoE, following the VR-CG QoE model in Section~\ref{sec:VR_CG_network}. Since frame rates are already determined in the first stage, this stage focuses on adaptive resolution selection for each virtual object in the scene. The objective function of the third-stage formulation is defined as:
\begin{equation}
    \maximize_{w_u^{r, o}} \sum_{u \in \mathcal{U}} \sum_{o \in \mathcal{O}_u} \left [ \lambda(u, o) \ln \left( \frac{\sum\limits_{r \in \mathcal{R}_u} w_u^{r, o}}{\min\limits_{r}(\mathcal{R}_u)} \right) \right ].
\end{equation}

Initially, we guarantee that each user receives data from the VR-CG game engine in at least one TTI, ensuring that every user is served and no user experiences service starvation. This requirement is expressed as:
\begin{equation}
    \sum_{k \in \mathcal{K}} r_u^{b, k} \geq 1, \qquad \forall u \in \mathcal{U}, b \in \mathbf{B}_u.
\end{equation}

Furthermore, we impose consistency constraints to ensure that if BS $b \in \mathbf{B}_u$ transmits data to user $u \in \mathcal{U}$ ($r_u^{b, k} = 1$), then a positive number of PRBs must be assigned to the user in the BS in this TTI ($n_u^{b, k} > 0$), constrained by the BS's total PRBs capacity. Conversely, if the BS $b \in \mathbf{B}_u$ is not transmitting ($r_u^{b, k} = 0$), no PRB must be allocated to the user $u \in \mathcal{U}$ from this BS in this TTI ($n_u^{b, k} = 0$). This restriction ensures a direct mapping between transmission decisions and resource allocation, preventing the accumulation of unassigned data in buffers across multiple TTIs. We formalize this as follows:
\begin{equation}
    b^{PRBs} r_u^{b, k} \geq n_u^{b, k}, \qquad \forall u \in \mathcal{U}, b \in \mathbf{B}_u, k \in \mathcal{K},
\end{equation}
\begin{equation}
    n_u^{b, k} \geq r_u^{b, k}, \qquad \ \ \ \forall u \in \mathcal{U}, b \in \mathbf{B}_u, k \in \mathcal{K}.
\end{equation}

Additionally, we define that the amount of PRBs allocated by each user during all TTIs is the same defined in the first stage solution. The following constraint defines this behavior:
\begin{equation}
    \sum_{k \in \mathcal{K}} n_u^{b, k} = y_u^b, \qquad \forall u \in \mathcal{U}, b \in \mathcal{B}.
\end{equation}

Next, we enforce that each user selects exactly one resolution for every virtual object within their FoV. This constraint is defined as follows:
\begin{equation}
    \sum_{r \in \mathcal{R}_u} w_u^{r, o} = 1, \qquad \forall u \in \mathcal{U}, o \in \mathcal{O}_u.
\end{equation}

To ensure that the newly selected resolutions for virtual objects remain feasible within the transmission resources allocated in the first stage, we must guarantee that the total traffic load generated in the third stage is at least equivalent to the load from the first stage. We calculate the total traffic load considering dynamic resolution per virtual objects as follows:
\begin{equation}
    Objects(u) = \sum_{r \in \mathcal{R}_u}\sum_{o \in \mathcal{O}_u} \Big ( \Phi(w_u^{r, o}) \eta\Big ) \Psi(\mathbf{z}_u),
\end{equation}
\noindent where $\Phi(w_u^{r, o})$ denotes the size of the object $o \in \mathcal{O}_u$ according to the resolution $r \in \mathcal{R}_u$. The parameter $\eta$ represents the compression rate of the encoding scheme, while $\Psi(\mathbf{z}_u)$ denotes the video frame rate selected for user $u \in \mathcal{U}$.

We need to ensure that the new object-level traffic load does not exceed the traffic load estimated in the first stage. The following constraint formalizes this requirement:
\begin{equation}
    Objects(u) \leq Load(u), \qquad \forall u \in \mathcal{U}.
\end{equation}

In the third stage, we ensure that each user's throughput demand, as defined in the first stage, is satisfied. However, since time is modeled in terms of TTIs in this stage, we need to compute the effective throughput offered to each user based on the number of TTIs during which they receive data from the BSs, as well as the number of PRBs allocated in each TTI. To estimate the total throughput for each user, we sum the throughput provided across all TTIs. The served throughput to user $u \in \mathcal{U}$ associated with BS $b \in \mathcal{B}_u$ in TTI $k \in \mathcal{K}$ is calculated as follows:
\begin{equation}
    TP(u, b, k) = n_u^{b, k} \log_2 \Big( 1 + SINR(u, b, k) \Big), 
\end{equation}
\noindent where $SINR(u, b, k)$ denotes the SINR of user $u \in \mathcal{U}$ for BS $b \in \mathcal{B}_u$ in TTI $k \in \mathcal{K}$, following Equation (\ref{eq:shannon}).

Finally, we must guarantee that the total throughput provided by all associated BSs during all TTIs meets the user's throughput demand. To formalize this requirement, we introduce the following constraint:
\begin{equation}
    Objects(u) \leq \sum_{b \in \mathbf{B}_u} \sum_{k \in \mathcal{K}} TP(u, b, k), \ \forall u \in \mathcal{U}.
\end{equation}

In VR-CG applications, synchronization between video frame generation and wireless resource allocation critically influences Motion-to-Photon (MTP) latency. Since VR-CG engines generate frames at a constant rate, the timing of PRB assignments directly affects the freshness of received frames. If a user's PRBs are allocated only at the beginning of the time window, the delivered frames will be among the earliest generated, leading to potentially outdated visual content. Conversely, if allocations occur only at the end, frames may span a broader range, introducing inconsistencies in MTP latency and disrupting synchronization between user actions and displayed content. To mitigate this, PRB allocations must be distributed evenly over time, allowing users to receive frames progressively throughout the time window.

We enforce this balanced allocation by grouping TTIs. We define a set of TTI groups as $\mathcal{T}$ where each $T \in \mathcal{T}$ corresponds to a group of TTIs $k \in \mathcal{K}$, where $T \subset \mathcal{K}$. In this way, we enforce that the users may receive transmission resources in all TTI groups with the following constraint:
\begin{equation}
\label{eq:TTI_groups}
    \sum_{k \in T} r_u^{b, k} \geq 1, \qquad \forall u \in \mathcal{U}, b \in \mathbf{B}_u, T \in \mathcal{T}.
\end{equation}

By solving the third stage subproblem, we define the resource allocation for VR-CG applications by dynamically adjusting the resolution of virtual objects based on user attention level. This stage ensures that users' throughput demands, as well as the QoE requirements established in the first stage, are met. Additionally, the fine-grained resolution selection and time-based resource management optimize network efficiency, reducing the impact of high resource usage associated with VR-CG applications. This enables the system to better handle varying network conditions and provide an enhanced immersive experience for users while minimizing bandwidth consumption.
\section{Scalable Resource Allocation for VR-CG}
\label{sec:VR-CG_solutions}

As discussed in \cite{Almeida:2025}, the joint optimization of resource allocation for VR-CG is an NP-hard problem, posing significant challenges for practical deployment in environments with hundreds or thousands of users. For example, the experiments presented in \cite{Almeida:2025} are limited to only 16 users, which falls short of the highly dense user scenarios anticipated in 6G networks. To address this limitation, we adopt a multi-stage optimization framework in this work, in which each stage represents a subproblem of the joint VR-CG resource allocation problem. This decomposition reduces the overall computational complexity by isolating decisions into smaller, more manageable components.

However, despite the advantages of this multi-stage formulation, each stage remains NP-hard \cite{Morais:2020, Liu:2016,Campos:2024}. Consequently, scalability issues persist when attempting to solve it optimally. To overcome this, we propose a set of scalable heuristic algorithms, one for each stage described in Section~\ref{sec:system_model}. These algorithms are designed to efficiently compute high-quality solutions within practical time constraints, making them suitable for real-world adoption.

\subsection{VR Experience-aware Algorithm}

To solve the problem formulated in the first stage, we propose the VR EXperience-aware Algorithm (VEXA), a many-to-many heuristic enhanced with a reorchestration module. VEXA performs user association based on wireless channel conditions, leveraging users' channel quality to guide the initial allocation. To avoid suboptimal greedy decisions, VEXA incorporates a reorchestration module that revisits previous associations based on users' QoE preferences and the current network state, enabling dynamic reassignment when needed. VEXA promotes fair resource distribution by explicitly considering the proposed QoE model (Equation~(\ref{eq:QoE_model})) into decision-making. The detailed functionality of VEXA is outlined in Algorithm~\ref{alg:VEXA}.

\begin{algorithm}[t]
\footnotesize
    \caption{VEXA heuristic algorithm}
    \label{alg:VEXA}
    \KwIn{$\mathcal{U}$, $\mathcal{B}$, $\mathcal{R}_u$, $\mathcal{F}_u$, SINR, $P(u)$ and $N$.} 
    \KwOut{User association and PRB allocation.}
    $E \gets \emptyset;$ \\
    \For{$u \in \mathcal{U}$}{
        $P^{BS}_u \gets$ list of BSs sorted by SINR($u$, $b$) value$;$\\
        $\textbf{C}_u \gets 0;$ $\textbf{R}_u \gets min(\mathcal{R}_u);$ $\textbf{F}_u \gets min(\mathcal{F}_u);$ \\
        \For{$b \in \mathcal{B}$}{
            Associated[$b$][$u$] $\gets \textbf{False};$ \\
            $\textbf{A}_b \gets b^{PRBs};$ $R[b][u] \gets 0;$
        }
    }
    \While{$|E| < |\mathcal{U}|$}{
    $u \gets$ random user $u \notin E;$\\
    \For{$b \in P^{BS}_u$}{
        \If{$\textbf{C}_u < N$}{
            $r^{PRB}_u\gets \ceil*{\frac{Load(u, b, \textbf{R}_u, \textbf{F}_u)}{\omega(b)}};$ \\
            \If{$\textbf{A}_b \geq \frac{r^{PRB}_u}{\textbf{C}_u + 1}$}{
                Associated[$b$][$u$] $\gets \textbf{True};$ \\
                $R[b][u] \gets \frac{r^{PRB}_u}{\textbf{C}_u + 1};$ \\
                $\textbf{C}_u \gets \textbf{C}_u + 1;$
                $E$.push($u$); \\
                $\textbf{A} \gets$ update BSs resources$;$
                \textbf{break}; 
            }
            \Else{
                $replaced \gets \textbf{False};$ \\
                \For{$u' \in$ \textnormal{Associated[$b$]}}{
                    \If{$P^{BS}_u > P^{BS}_{u'}$}{
                        Associated[$b$][$u'$] $\gets \textbf{False};$ \\
                        $\textbf{C}_{u'} \gets \textbf{C}_{u'} - 1;$
                        $E$.pop($u'$); \\
                        $\textbf{A} \gets$ update BSs resources$;$\\
                        $replaced \gets \textbf{True};$ \\
                        \textbf{break}; 
                    }
                }
                \If{$replaced$}{
                    \textbf{continue}; 
                }
            }
        }
        \Else{
            \textbf{break}; 
        }
    }
}
    $\text{MaximizeQoE(Associated}, \mathcal{U}, \mathbf{R}_u, \mathbf{F}_u, \mathcal{R}_u, \mathcal{F}_u, \textbf{A});$
\end{algorithm}

The VEXA algorithm takes as input the same parameters defined in the first-stage problem formulation. It operates as a centralized application that manages user association across multiple BSs. The algorithm begins by initializing an empty list $E$, which stores the set of admitted users, and a matrix $R$ for PRB allocation. During initialization, each user is assigned their minimum supported resolution and frame rate to ensure baseline QoE. The core of the algorithm, detailed in lines 8 -- 26, iteratively attempts to associate users with BSs based on preferences computed from SINR values. If a preferred BS has sufficient capacity, the user is admitted; otherwise, the algorithm evaluates the BS's currently admitted users and may replace lower-priority users with higher-priority ones. Users that cannot be admitted move to their next preferred BS. Finally, in line 27, the VEXA application invokes the QoE maximization procedure described in Algorithm~\ref{alg:QoE_maximization}. This algorithm provides near-real-time control over BSs regarding user association, handovers, and PRB allocation, and can be implemented in the Open-RAN (O-RAN) ecosystem, for example, as an xApp.

The QoE maximization process, in Algorithm~\ref{alg:QoE_maximization}, starts from the previously computed user association. It then executes an iterative procedure enhancing users' QoE based on a utility function, defined as the difference between each user's maximum achievable QoE and their current QoE. This metric prioritizes users whose perceived QoE can increase the most. Thanks to the Weber-Fechner law underlying our QoE model, which states that perceived satisfaction grows logarithmically with stimulus intensity, users with lower QoE experience larger perceived gains for the same resource allocation. As a result, the algorithm naturally favors boosting under-served users first, leading to a balanced and fair distribution of resources across users.

\begin{algorithm}[t]
    \caption{MaximizeQoE()}
    \label{alg:QoE_maximization}
    \KwIn{Associated, $\mathcal{U}$, $\mathbf{R}_u$, $\mathbf{F}_u$, $\mathcal{R}_u$, $\mathcal{F}_u$ and $\textbf{A}$.}
    \KwOut{Image resolution and frame rate selection.}
    \For{$u \in \mathcal{U}$}{
        Util($u$) $\gets$ QoE$((max(\mathcal{R}_u) - \textbf{R}_u) + (max(\mathcal{F}_u) - \textbf{F}_u));$
    }
    $changed \gets \textbf{True};$ \\
    \While{changed}{
        $changed \gets \textbf{False};$ \\
        $\textbf{U} \gets $ list of users sorted by Util($u$) value$;$\\
        \For{$u \in \textbf{U}$}{
            \eIf{P(u) = 1}{
                \For{$r \in \mathcal{R}_u$}{
                    \If{$feasible(\textbf{A}, u, r);$}{
                        $\textbf{R}_u \gets r;$ 
                        $changed \gets \textbf{True};$ \\
                        $\textbf{A} \gets$ update BSs  resources$;$\\
                        \textbf{break}$;$
                    }
                }
            }
            {
                \For{$f \in \mathcal{F}_u$}{
                    \If{$feasible(\textbf{A}, u, f);$}{
                        $\textbf{F}_u \gets f;$ 
                        $changed \gets \textbf{True};$ \\
                        $\textbf{A} \gets$ update BSs  resources$;$\\
                        \textbf{break}$;$
                    }
                }
            }
        }
    }
    \textbf{return} Associated, $\mathbf{R}_u, \mathbf{F}_u$$;$
\end{algorithm}

The VEXA algorithm addresses the first stage problem formulation by determining user admission, communication resource allocation, and maximizing users' QoE. The overall computational complexity of VEXA is $\mathcal{O}(|\mathcal{U}|^2|\mathcal{B}|^2 + |\mathcal{U}||\mathcal{R}||\mathcal{F}|)$. However, since the number of available resolutions and frame rates per user is typically much smaller than the number of users and BSs in the network, the practical complexity is dominated by $\mathcal{O}(|\mathcal{U}|^2|\mathcal{B}|^2)$. Moreover, the algorithm can be efficiently parallelized in real deployments by clustering users whose resource allocations do not interfere with one another, further improving scalability.

\subsection{Game Engine Placement and Adaptive Routing}

To address the problem formulated in the second stage, we propose the Game Engine Placement and Adaptive Routing (GEPAR) heuristic. GEPAR follows a greedy placement strategy with principles from First-Fit Decreasing. The algorithm selects the game engine's placement based on the previously established user–BS associations, considering both computing and communication requirements. It constructs a preference list of CNs for each user, ranked according to latency and operational cost. To reduce deployment costs, GEPAR prioritizes the reuse of already active CNs whenever possible. Throughout the process, the heuristic ensures that all computing capacity and communication constraints are satisfied. The GEPAR heuristic algorithm is defined in Algorithm \ref{alg:GEPAR}.

\begin{algorithm}[tt]
\footnotesize
    \caption{GEPAR heuristic algorithm}
    \label{alg:GEPAR}
    \KwIn{$\mathcal{U}$, $\mathbf{B}$, $\mathcal{C}$, $\mathcal{P}_b^c$, $\mathbf{v}^F_c$, $\mathbf{v}^V_c$, $\mathbf{G}(u)$, $\mathbf{C}(u)$ $\mathbf{M}(u)$ and $\mathbf{N}(u)$.}
    \KwOut{VR-CG game engine placement for all users.}
    $Active = \emptyset;$ \\
    \For{$u \in \mathcal{U}$}{
        $\mathbf{P}^{\mathcal{C}}_u \gets$ list of feasible CNs sorted by cost$;$\\
        $\mathbf{S}_u \gets \emptyset;$\\
    }
    \For{$u \in \mathcal{U}$}{
        $Tested \gets \emptyset;$
        \textit{Placed} $\gets \textbf{False};$ \\
        \While{\textbf{not} Placed}{
            $c \gets \mathbf{P}^{\mathcal{C}}_u$.pop$;$\\
            \If{$c \notin Active$}{
                \For{$n \in Active$}{
                    \If{$\mathbf{v}^F_c + \mathbf{v}^V_c \geq \mathbf{v}^V_{n}$ \textbf{and} $n \notin Tested$ \textbf{and} $n \in \mathbf{P}_u^{\mathcal{C}}$}{
                        $c \gets n;$
                    }
                }
            }
        $Tested \gets c;$ \\
        \If{$Available(c^{GPU}, c^{CPU}, c^{RAM}, c^{NET})$}{
            \For{$p \in$ k-shortest$(\mathcal{P}^c_b)$}{
                \If{$p$ has enough capacity}{
                    $Placed \gets \mathbf{True};$
                    $Active \gets c;$ \\
                    $\mathbf{S}_u \gets c;$ \\
                    $Update(c^{GPU}, c^{CPU}, c^{RAM}, c^{NET}, p);$
                }
            }
        }
        }
    }
    \textbf{return} $\mathbf{S}_u$
\end{algorithm}

The GEPAR algorithm takes as input the solution from the first stage, comprising user association and communication resource allocation, along with the parameters defined in the second-stage problem formulation, such as the computing capacities of CNs, routing paths, and link capacities. The algorithm begins by initializing an empty list of active CNs. It then computes each user's preference list $\mathbf{P}_u^{\mathcal{C}}$, which ranks CNs by operational cost, considering only those with feasible latency. Starting at line 5, the algorithm enters an iterative process to determine the VR-CG game engine placement for all users. For each user, the algorithm selects the CN with higher preference and check its feasibility. In line 9, the algorithm checks whether the selected CN is already active. If not, it attempts to reuse a previously activated node to reduce operational cost. Once a target CN is selected, at line $14$, the algorithm verifies whether it has sufficient computing capacity and routing bandwidth to host the user's VR-CG game engine. If feasible, the placement is selected and the corresponding capacities are updated. The algorithm terminates once all users have had their game engines successfully placed.

The GEPAR heuristic algorithm addresses the second-stage problem formulation by determining the placement of VR-CG game engines for all users admitted to the network. The overall computational complexity of GEPAR is $\mathcal{O}(|\mathcal{U}||\mathcal{C}|k)$, where $|\mathcal{U}|$ is the number of users, $|\mathcal{C}|$ is the number of CNs, and $k$ is the number of candidate paths considered by the $k$-shortest path algorithm. This complexity can be mitigated through parallelization, as each BS can independently compute the game engine placement for its associated users, enabling highly scalable execution in practical scenarios.

\subsection{Attention-aware MTP-based PRBs scheduling}

To address the problem formulated in the third stage, we propose the Attention-aware MTP-based PRB Scheduling (AMPS) algorithm, a heuristic for attention and MTP-aware resource allocation PRB scheduling in NG-RAN. This algorithm refines the selection of resolution and frame rate for each user based on their attention level to virtual objects, improving the baseline solutions from the previous stages. It iteratively enhances the resolution of high-attention objects while respecting the computing and communication constraints. Furthermore, the AMPS algorithm performs PRB scheduling by mapping the data transmission to specific TTIs within the scheduling window. The scheduling process is guided by an MTP-aware strategy that aims to maximize user QoE by minimizing MTP latency and ensuring the timely delivery of visual content aligned with user attention. The AMPS heuristic is presented in Algorithm \ref{alg:AMPS}.

\begin{algorithm}[t]
\footnotesize
    \caption{AMPS heuristic algorithm}
    \label{alg:AMPS}
    \KwIn{$\mathcal{U}$, $\mathcal{O}_u$ $\mathbf{B}_u$, $\mathcal{K}$, $\mathcal{T}$.}
    \KwOut{Object resolution for all users.}
    $\mathcal{U}^{max}_{res} \gets \mathcal{U}$ sorted by maximum resolution supported$;$\\
    \For{$u \in \mathcal{U}$}{
        $\mathcal{O}^{att.}_u \gets \mathcal{O}_u$ sorted by attention level$;$\\
        $\mathcal{O}^{leng.}_u \gets \mathcal{O}_u$ sorted by length$;$\\
        \For{$o \in \mathcal{O}^{att.}_u$}{
            $\mathbf{R}_u^o \gets$ Resolution defined in the first stage$;$
        }
    }
    \For{$u \in \mathcal{U}^{max}_{res}$}{
        \For{$o \in \mathcal{O}_u^{att.}$}{
            \eIf{$\mathbf{R}^o_u$ has max. resolution}{
                \textbf{continue}$;$
            }
            {
                $r_u^o \gets$ higher target resolution \\
                \eIf{$r_u^o$ is feasible}{
                    $\mathbf{R}^o_u \gets r_u^o ;$ \\
                    $\mathbf{y}_u^b \gets$ PRB allocation$;$\\
                    Update resource usage$;$\\
                }
                {
                    \For{$p \in \mathcal{O}_u^{leng.}$}{
                        \If{$\lambda(u, o) > \lambda(u, p)$}{
                            $r_u^{p} \gets$ lower target resolution$;$\\
                            \If{$r_u^{p}$ \textbf{and} $r_u^o$ is feasible}{
                                $\mathbf{R}^{p}_u \gets r_u^{p};$
                                $\mathbf{R}^o_u \gets r_u^o ;$ \\
                                $y_u^b \gets$ PRB allocation$;$\\
                                Update resource usage$;$\\
                            }
                        }
                    }
                }
            }
        }
    }
MTPsched($\mathbf{R}_u^o$ $\mathcal{U}$, $\mathbf{B}_u$, $y_u^b$ $\mathcal{K}$, $\mathcal{T}$);
\end{algorithm}

The AMPS algorithm begins by sorting users based on the resolution capacity of their HMDs. For each user, it identifies the virtual objects within their FoV, ordering them by attention level and visual prominence. Initially, the resolution of each object is set according to the first stage solution. Starting at line 7, the algorithm iterates over each user's object list and attempts to increase the resolution of each object. If the upgrade is feasible under the current computing and communication constraints, the higher resolution is assigned, and the resource usage is updated. If it is not feasible, the algorithm attempts to reallocate resources by lowering the resolution of lower-attention objects. After processing all users and their corresponding objects, the algorithm proceeds to assign the PRBs and then calls the MTPsched algorithm. The MTPsched heuristic is described in Algorithm \ref{alg:MTPsched}.

\begin{algorithm}[t]
\footnotesize
    \caption{MTPsched algorithm}
    \label{alg:MTPsched}
    \KwIn{$\mathcal{U}$, $\mathbf{B}_u$, $y_u^b$ $\mathcal{K}$ and $\mathcal{T}$.}
    \KwOut{PRB scheduling during all TTIs.}
    \For{$b \in \mathbf{B}_u$}{
        \For{$k \in \mathcal{K}$}{
            $transmit[b][k] \gets \emptyset;$
        }
        \For{$u \in \mathcal{U}$}{
            $\mathbf{y}_u^b \gets 0;$\\
        }
    }
    $finished \gets \textbf{False};$ \\
    \While{\textbf{not} finished}{
        $u \gets$ user not finished from $\mathcal{U};$\\
        \For{$b \in \mathbf{B}_u$}{
            $\mathbf{k} \gets max(|\mathcal{T}|, y_u^b - \mathbf{y}_u^b)$ \\
            \While{$i \leq \mathbf{k}$}{
                $\mathbf{T} \gets \frac{i(|\mathcal{T}| + 1)}{(\mathbf{k} + 1)};$
                $i \gets i + 1;$
            }
            \For{$k \in \mathbf{k}$}{
                \eIf{$transmit[b][k].size < b^{PRBs}$}{
                    $transmit[b][k].push(u);$ \\
                    $\mathbf{y}_u^b \gets \mathbf{y}_u^b + 1;$
                }
                {
                    Get the closest TTI $k^*$ available$;$ \\
                    $transmit[b][k^*].push(u);$ \\
                    $\mathbf{y}_u^b \gets \mathbf{y}_u^b + 1;$
                }                
            }
        }
        \If{$\mathbf{y}_u^b = y_u^b, \ \forall u \in \mathcal{U}$}{
            $finished \gets \textbf{True};$
        }
    }
\textbf{return} $\mathbf{R}_u^o$, $transmit;$
\end{algorithm}

The MTPsched algorithm initialize an empty matrix, in line 3, that represents the BS spectrum, storing the set of users transmitting in each TTI. In line 5, it initializes the number of received PRBs for each user to zero. The iterative scheduling process starts in line 7, where the algorithm allocates PRBs to users. For each user that still requires PRBs the algorithm iterates over all BSs serving that user. In line 10, it computes the number of PRBs to allocate in the current iteration, distributing them across TTI groups as defined in Section~\ref{subsec:third_stage}. This strategy enables MTPsched to reduce MTP latency by spreading the allocation across different time windows. The PRB selection process begins in line 14. If the preselected TTI has available PRBs, they are directly assigned to the user. If not, in line 18, the algorithm searches for the closest TTI with available PRBs to allocate. This process repeats until all users receive the number of PRBs defined in the first-stage solution, ensuring the data required for each user is transmitted within the throughput and latency constraints.

The overall computational complexity of the AMPS heuristic is given by $\mathcal{O}(|\mathcal{U}|^2|\mathcal{O}||\mathcal{B}| + |\mathcal{K}|^2|\mathcal{U}||\mathcal{B}|)$, where the first term corresponds to the resolution adaptation process and the second term represents the complexity of the MTPsched scheduling algorithm. In practical scenarios, the number of TTIs (determined by the 5G/6G numerology) typically exceeds the number of users associated with each BSs. As a result, the complexity of AMPS is dominated by the MTPsched component, i.e., $\mathcal{O}(|\mathcal{K}|^2|\mathcal{U}||\mathcal{B}|)$. To improve scalability in realistic deployments, the PRB allocation can be performed independently at each BSs. This parallelization strategy significantly reduces computation time, as MTPsched can be executed locally per BS, rather than jointly for the entire network.
\section{Evaluation}
\label{sec:evaluation}

This section presents our evaluation setup, which is composed of two datasets we collected from \textit{Steam}\footnote{https://store.steampowered.com/}, the largest gaming platform with an estimated 75\% of the market share \cite{toy2018large}. Next, we detail the obtained results evaluating the proposed multi-stage optimization framework and the scalable heuristics in practical scenarios.

\subsection{Setup}
\label{subsec:experimental_setup}

We collected two datasets, one describing VR headsets and the other detailing VR games. Both datasets are used as input for VEXA, and bring our evaluation scenarios closer to what is expected in terms of user hardware (VR headsets) and user preferences (VR games).

\textbf{VR headsets data} -- This dataset details hardware specifications of the most common VR headsets. To generate this dataset, we relied on the \textit{Steam Hardware \& Software Survey}\footnote{https://store.steampowered.com/hwsurvey/Steam-Hardware-Software-Survey}, specifically the January 2025 version, which describes not only the most common VR headsets among their users, but also the user percentage for each headset.  Next, we collected information regarding \textit{per-eye resolution} and \textit{refresh rate} for each device. To achieve that, for each headset in the \textit{Steam Hardware \& Software Survey}, we searched on \textit{VRcompare}\footnote{https://vr-compare.com/} the required information. Finally, our dataset is composed by a variety of VR headsets with information regarding their market share, i.e., user percentage for each device (extracted from \textit{Steam}), and their hardware specifications, i.e., \textit{per-eye resolution} and \textit{refresh rate} (extracted from \textit{VRcompare}). In short, the dataset contains 22 unique VR headsets, with 15 unique \textit{per-eye resolutions} (varying from 960×1080p up to 2880×2720p) and 6 unique \textit{refresh rates} (varying from 72 Hz up to 144 Hz).

\textbf{VR games data} -- This dataset details for VEXA relevant software (game) information of the most common VR games on \textit{Steam}. For this dataset, we relied on \textit{SteamSpy}\footnote{https://steamspy.com/}, an API that enabled us to select only \textit{Steam} games tagged as VR-capable. With the list of available VR games on \textit{Steam}, we collected a variety of relevant information, including: \textit{ID}; \textit{Name}; \textit{Developer}; \textit{Publisher}; \textit{Owners} (number of users who own this game on Steam); \textit{Average playtime} (both average user playtime since March 2009 and in the last two weeks, in minutes); Median playtime (both median user playtime since March 2009 and in the last two weeks, in minutes); \textit{CCU} (Yesterday's peak concurrent users) and \textit{Price} (current US price). We also collected tag information if the game is considered \textit{Competitive}, i.e., a game in which the users compete against each other, or not.

We extend our dataset with more relevant information that were not available on \textit{SteamSpy}. To achieve that, we relied on \textit{Steam Charts}\footnote{https://steamcharts.com/} to obtain the average number of players in the last 30 days for each VR game. With this information, we calculated each user percentage for each VR game, based on the average number of users that played this game in the last 30 days, and the sum of this average for every game. This empowered our data with the average number of players and the user percentage per game in the last 30 days. In short, our final dataset contains the aforementioned information of 1,247 games from 1,072 developers.

In the evaluation, we adopt the simulation setup proposed in \cite{Esper:2023}. The scenario considers a 6G network deployed over a square area of 4 km$^2$ (2 km × 2 km), with user mobility simulated for one hour using the \textit{Simulation of Urban MObility}\footnote{https://eclipse.dev/sumo/} (SUMO) framework (version 1.25.0). A total of 1,250 users are distributed throughout the area, moving by car, bus, or remaining stationary, reflecting the assumption that pedestrians are unlikely to engage in VR-CG applications. All users request access to VR-CG services and demand allocation of communication and computing resources according to their channel conditions. The optimization models were solved using CPLEX (version 22.11), capable of handling both linear and non-linear constraints, with Python (version 3.10.12) using the docplex (version 2.27.239).

The network includes 10 BSs, each colocated with a CN, and configured with a coverage radius of 400 meters. CNs are interconnected through a ring-based crosshaul topology \cite{Morais:2020}. Each BS operates with 20 MHz of bandwidth in distinct sub-6 GHz frequency channels, adopting numerology 1 as specified by 3GPP \cite{TS38211:2025}. Accordingly, each BS supports TTIs of $0.5$ ms, a subcarrier spacing of 30 kHz, and 12 subcarriers per PRB, resulting in 360 kHz per PRB and a total of 56 PRBs. In addition to the VR-CG transmission load, we incorporate background traffic occupying 70\% of the available bandwidth, following the practical traffic described in \cite{Esper:2023}, to reflect realistic network resource competition. During the evaluation, we considered a simulation window of 100 seconds, divided into 1-second intervals. At each interval, the network state is updated and the optimization models and heuristic algorithms are applied to compute resource allocation decisions. This setup allows us to capture dynamic variations in user demand and network load while evaluating the performance of the proposed solutions over time.

\subsection{Results}
\label{subsec:evaluation}

Following the structure of our optimization framework, we organize the analysis into three categories: (i) user association and PRB allocation, (ii) game engine placement and routing, and (iii) attention-aware resource scheduling. For each category, we present optimal and heuristic solutions produced by our approaches and evaluate their effectiveness by comparing them against relevant baseline approaches from the literature.

\subsubsection{User association and PRB allocation}
\label{subsubsec:first_stage_solutions}

In this section, we evaluate our heuristic algorithm, VEXA, by comparing its results with those of the optimal model and two baseline approaches from the literature. The first baseline is Single Association (SA), in which each user is connected to exactly one BS. This reflects the native association mechanism in 5G networks, where each UE is typically served by a single BS unless explicit multi-connectivity features are enabled. The second baseline is Dual Connectivity (DC), a Multi-Connectivity (MC) mechanism defined by 3GPP \cite{TS36300:2025} that allows each user to be simultaneously served by up to two BSs.

\begin{figure}[t]
    \centering
    \includegraphics[width=1\linewidth]{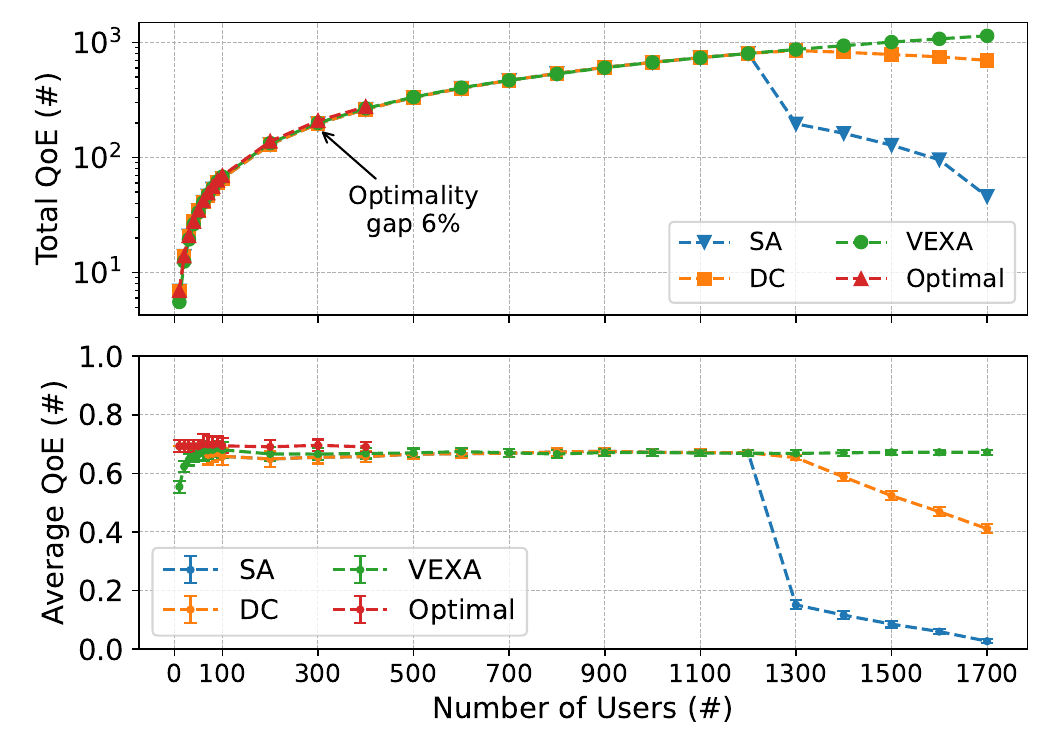}
    \caption{Total and average QoE comparison.}
    \label{fig:total_and_avg_QoE}
\end{figure}

Figure~\ref{fig:total_and_avg_QoE} presents the total and average QoE achieved by each model as a function of the number of users in the system. Although the optimization model delivers the highest QoE values, it is only capable of solving instances with up to 400 users due to its high computational complexity. We allowed the solver to run for up to 24 hours, but it was unable to find feasible solutions for larger instances within this time limit. On the other hand, the QoE values obtained by the VEXA heuristic are close to those of the optimal model. The maximum optimality gap was 6\% in the total QoE, while the average gap across all instances was 1.34\%. These results demonstrate the effectiveness of VEXA in providing near-optimal solutions.

In general, the heuristic approaches are capable of solving larger instances, maintaining QoE levels close to those achieved by the optimal solution. Notably, VEXA outperforms the other heuristics as the number of users increases, delivering higher QoE values. This is largely due to VEXA's multi-connectivity support, which proves particularly beneficial in high-density scenarios by mitigating the effects of resource scarcity.

\begin{figure}[t]
    \centering
    \includegraphics[width=1\linewidth]{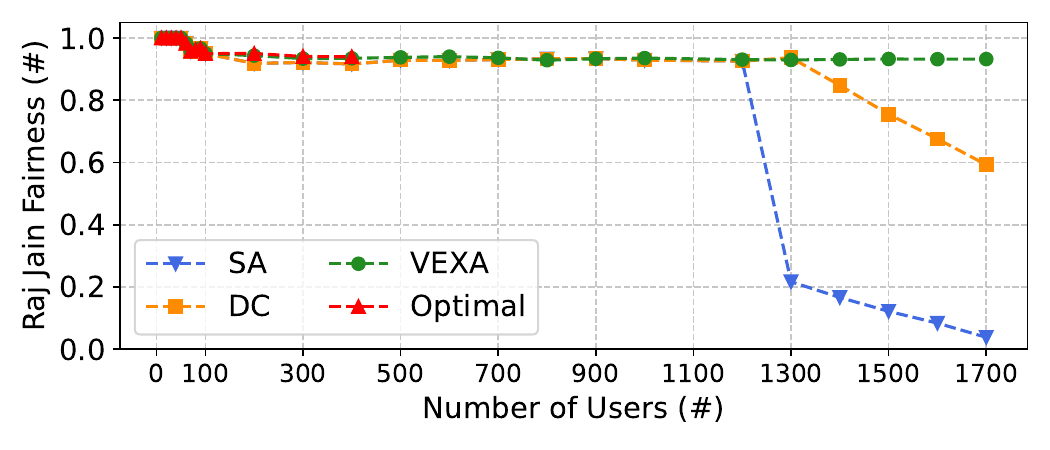}
    \caption{Raj Jain fairness index comparison.}
    \label{fig:raj-jain-comparison}
\end{figure}

To evaluate the fairness of QoE allocation among users in each model, we compute the Raj Jain’s fairness index, a widely used metric for assessing fairness in multi-user resource allocation scenarios \cite{jain1984quantitative}. This index ranges from 0 to 1, where values closer to 1 indicate higher fairness in resource distribution. Figure~\ref{fig:raj-jain-comparison} presents the fairness index for each model as a function of the number of users. As expected, the optimization model achieves the highest fairness, but it is limited to 400 users due to its computational complexity. In contrast, VEXA consistently maintains a high fairness level even as the number of users increases, demonstrating its scalability and robustness in dense network conditions. This result indicates that VEXA not only provides high QoE in large-scale scenarios but also ensures equitable resource distribution, a benefit of its QoE model grounded in the Weber–Fechner satisfaction law.

\begin{figure}[t]
    \centering
    \includegraphics[width=1\linewidth]{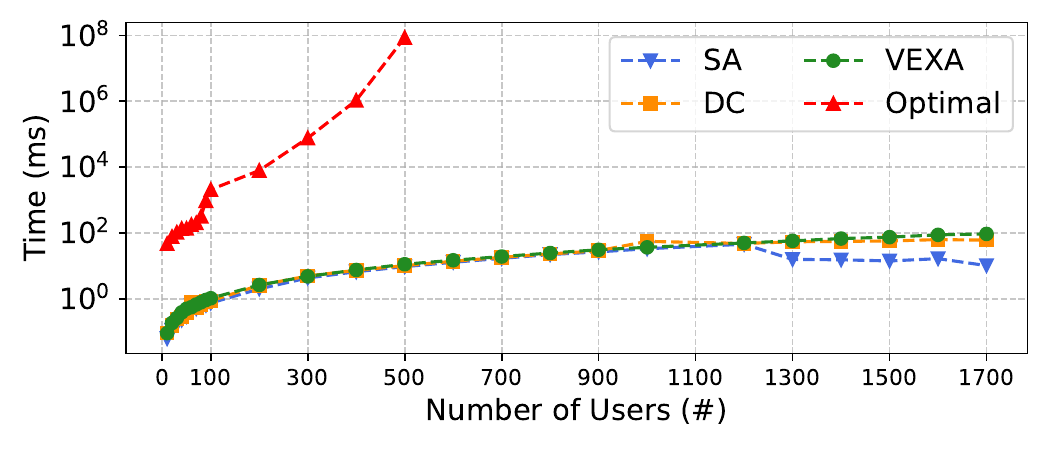}
    \caption{Computational time for each model.}
    \label{fig:time_comparison}
\end{figure}

In Figure \ref{fig:time_comparison}, we present the computational time required by each solution to solve different instances. The optimal model exhibits higher computational complexity, requiring more time in all scenarios. With 500 users, the solver reaches the imposed 24-hour runtime limit and terminates the search without finding an optimal solution. In contrast, the heuristic approaches demonstrate greater scalability, solving all tested instances in significantly less time. In particular, VEXA completes the computation for 1,200 users in under 0.1 seconds, demonstrating its ability to solve the problem in practical scenarios.

\begin{figure}[t]
    \centering
    \includegraphics[width=1\linewidth]{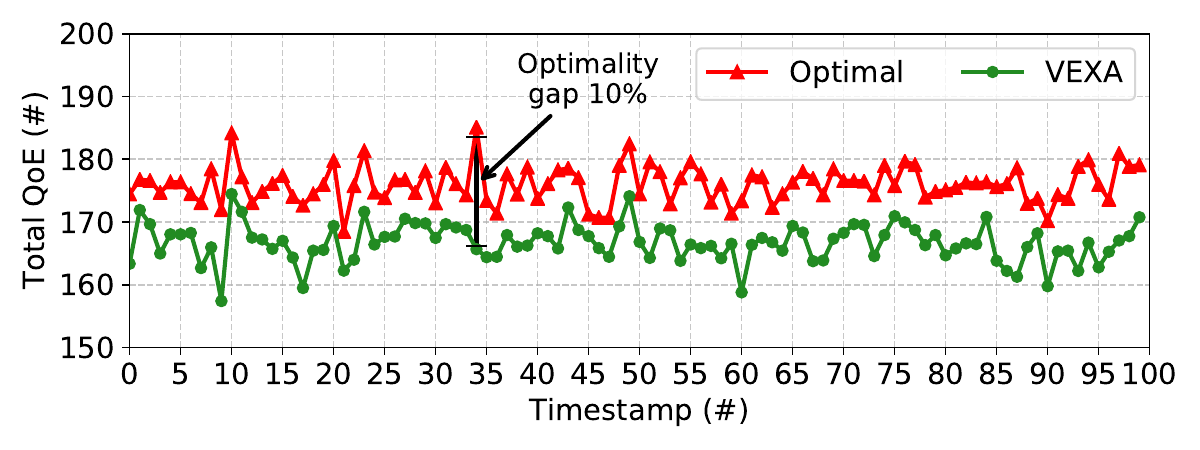}
    \caption{Users QoE variation over time.}
    \label{fig:QoE_variation}
\end{figure}

Finally, in Figure~\ref{fig:QoE_variation}, we evaluate the adaptability of the proposed VEXA heuristic in comparison to the optimal model. We fixed the number of users at 400 and simulated 100 timestamps, each representing one second of network activity, including user mobility and video frame transmission. Although the optimal model consistently yields higher total QoE values, it requires approximately 16 minutes to solve each instance, which corresponds to just one second of real-time operation. In contrast, VEXA solves each instance in under 100 milliseconds. Additionally, VEXA closely tracks the QoE variation pattern of the optimal model. These results demonstrate that, while the heuristic delivers near-optimal QoE values, it also exhibits adaptability to network dynamic, achieving a maximum optimality gap of 10\% and an average gap of 5\%. 

\subsubsection{Game engine placement and routing}
\label{subsubsec:second_stage_solutions}

In this section, we evaluate the GEPAR heuristic, that addresses the VR-CG game engine placement considering multipath routing in the transport network. We compare the solutions produced by GEPAR with those obtained from the optimal model, as well as two baselines from the literature: (i) the Single Path Placement baseline, which restricts routing to a single path between base stations and computing nodes; and (ii) the Unconstrained Placement model \cite{Wang:2018}, which incorporates all placement constraints directly into the objective function rather than treating them as hard constraints.

\begin{figure}[t]
    \centering
    \includegraphics[width=1\linewidth]{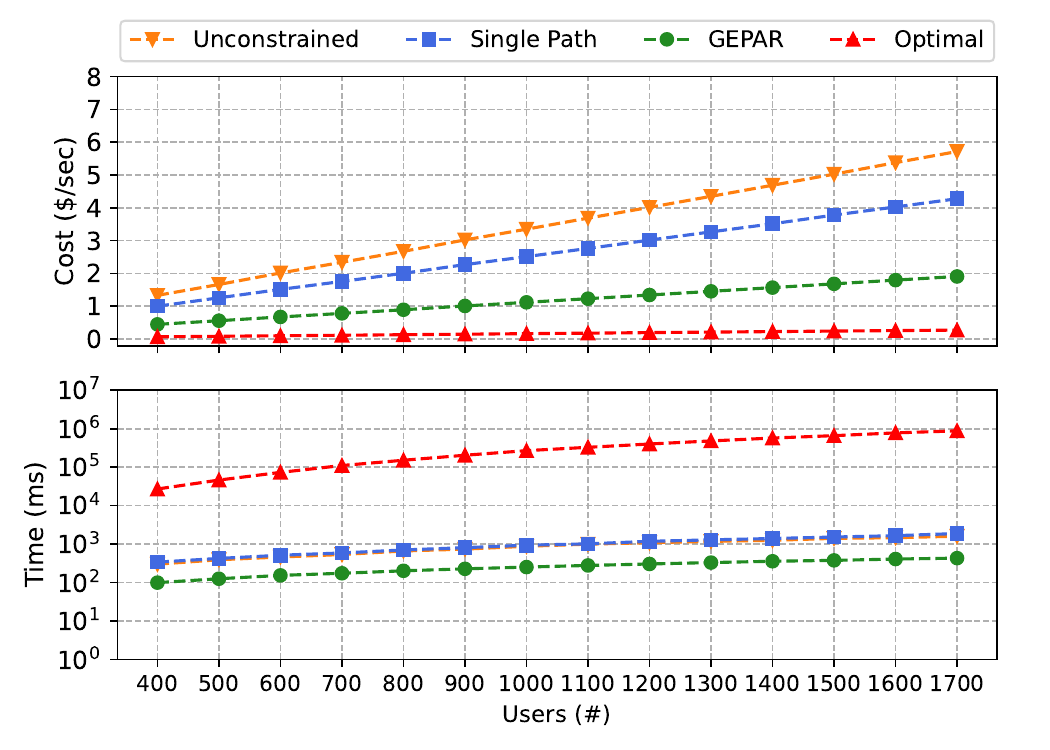}
    \caption{Comparing total cost and computational time.}
    \label{fig:cost_and_time}
\end{figure}

Figure~\ref{fig:cost_and_time} presents the results of the VR-CG game engine placement models for different numbers of users.

To compute the monetary cost associated with computing resource usage, we assume that the network operator outsources computational capacity from a cloud service provider. In particular, we consider Amazon Web Services (AWS), a widely adopted cloud platform that offers on-demand virtual machines and accelerator-equipped instances. For our evaluation, we use the pricing of an AWS instance equipped with an NVIDIA L4 GPU\footnote{https://aws-pricing.com/g6.48xlarge.html}, which is suitable for remote frame rendering and real-time video streaming. The cost is computed based on the instance usage across different deployment tiers (cloud, regional, and edge sites) reflecting realistic pricing variations across geographic and architectural layers.

For scenarios with a higher number of users, all models are initialized using the VEXA heuristic solutions. While the optimization model achieves the lowest total cost, our proposed GEPAR algorithm delivers near-optimal solutions with substantially reduced computational time. This performance gain is primarily due to the lower complexity of the GEPAR heuristic, in contrast to the Single Path and Unconstrained models, which rely on optimal search formulations. The results demonstrate that our multipath routing strategy leads to greater cost savings as user density increases, especially in scenarios where routing and computing resources are more heavily contested.

\begin{figure}[t]
    \centering
    \includegraphics[width=1\linewidth]{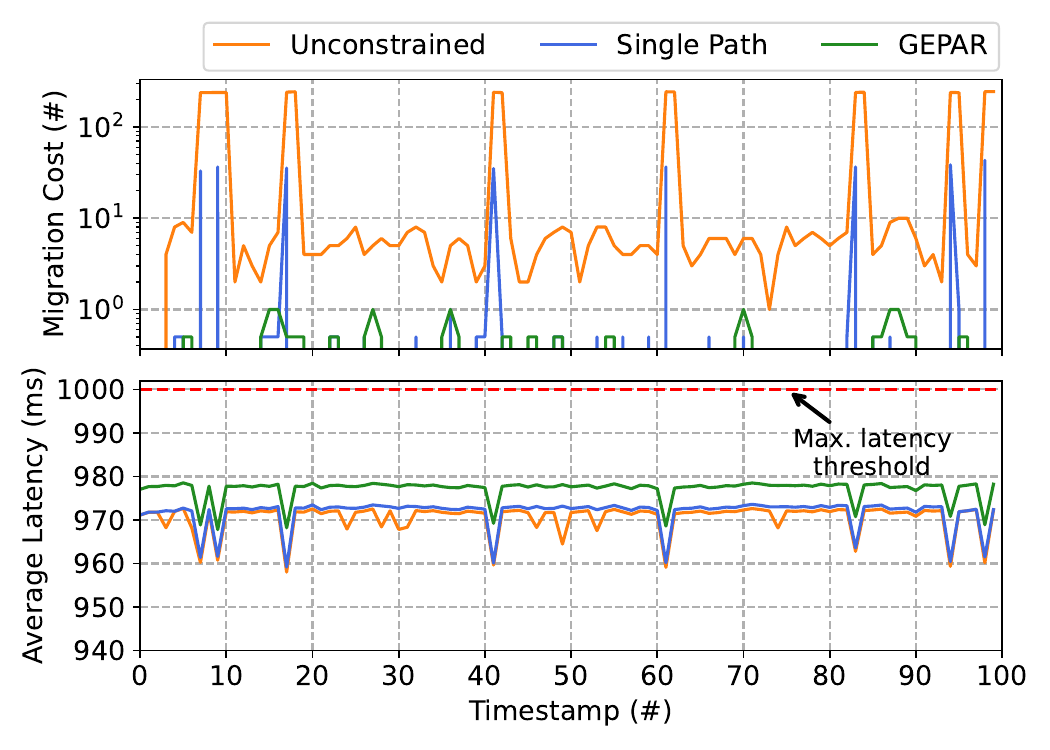}
    \caption{Migration cost and average latency comparison.}
    \label{fig:migration_and_latency}
\end{figure}

To assess the adaptability of the proposed GEPAR heuristic, we evaluate a scenario with 250 users over 100 timestamps. We compare its performance against the Unconstrained model and the Single Path model. The optimal solution, using multipath routing, is not included since it is unable to solve this instance with 250 users withing a 24 hours of run time limit.

Figure~\ref{fig:migration_and_latency} shows the migration cost and the average user-perceived latency for all approaches. The Unconstrained model exhibits the highest migration cost, highlighting that its objective function requires careful parameter tuning and is therefore less suitable for highly dynamic environments. The Single Path model also results in elevated migration costs because restricting the transport network to a single path often forces more frequent relocations of the VR-CG game engine to satisfy latency requirements. In contrast, the proposed GEPAR heuristic achieves substantially lower migration costs, reducing them by approximately 35\% on average. This improvement stems from its multipath routing strategy, which enables flexible traffic rerouting without relocating the game engine, thereby maintaining service continuity while minimizing migration overhead.

Regarding user-perceived latency, all approaches produced solutions with user latency below the maximum threshold. However, the Unconstrained model achieves lower latency values, as it is directly optimized in its objective function. The other approaches, including GEPAR, aim to satisfy the latency constraint without actively minimizing it.

\subsection{Attention-aware resource scheduling}
\label{subsubsec:third_stage_solutions}

In this section, we evaluate the effectiveness of the AMPS heuristic and the MTPsched algorithm. AMPS enhances user QoE by leveraging an attention-aware mechanism that prioritizes resource allocation based on users' focus on virtual objects. MTPsched, in turn, introduces an MTP-aware PRB scheduling strategy aimed at minimizing MTP latency. 

\begin{figure}[t]
    \centering
    \includegraphics[width=1\linewidth]{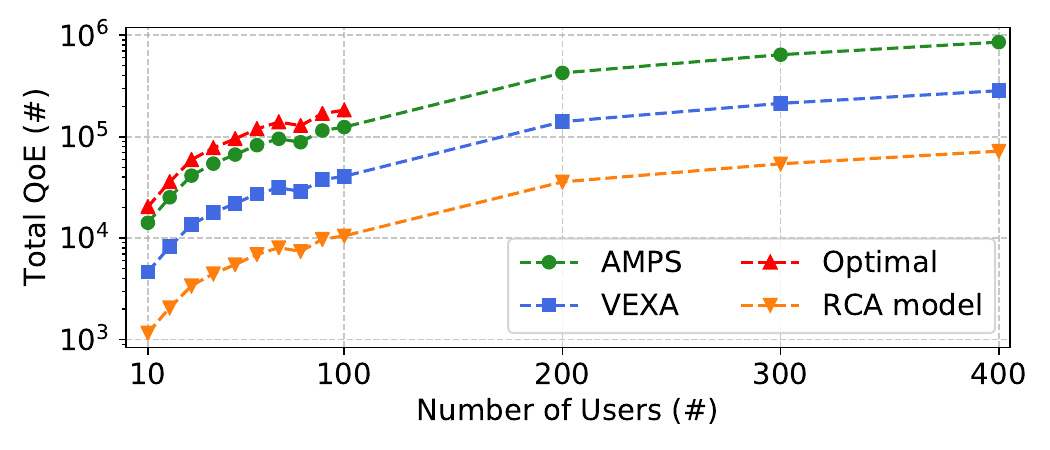}
    \caption{Comparing total QoE of VEXA, AMPS and RCA models.}
    \label{fig:comparing_1st_3rd_RCA}
\end{figure}

Figure~\ref{fig:comparing_1st_3rd_RCA} presents the total QoE comparison among four approaches: our VEXA heuristic (first-stage optimization), the RCA model from \cite{Du_2023}, our AMPS heuristic, and the optimal solution from the third stage. As before, we enforce a 24-hour maximum solving time for the optimization model, which allows the optimal solver to produce solutions only for scenarios with up to 100 users due to the high computational complexity of the third-stage formulation. Since the third stage focuses on resource allocation within each BS, we consider up to 400 users simultaneously associated with a BS. This setup allows us to evaluate the solutions across a spectrum of conditions, from low-demand scenarios to high-density deployments, such as crowded public spaces, where BSs must serve an unusually large number of users, thereby assessing the scalability of our solutions under extreme conditions.

As previously discussed, VEXA selects a baseline solution for each user in terms of image resolution and frame rate. This baseline acts as the starting point for the subsequent stages. In the third stage, AMPS builds upon this initial allocation by applying an attention-aware resolution refinement, thereby enhancing users' QoE by up to 50\% and presenting near-optimal solutions. As a result, AMPS consistently achieves higher total QoE values across all scenarios by increasing the resolution of objects based on user attention. Additionally, both VEXA and AMPS outperform the RCA optimization model. This improvement stems from our approach’s enhanced fairness in resource distribution, whereas the RCA model tends to favor users with better wireless channel quality \cite{Almeida:2025}.

\begin{figure} [t]
    \centering
    \includegraphics[width=1\linewidth]{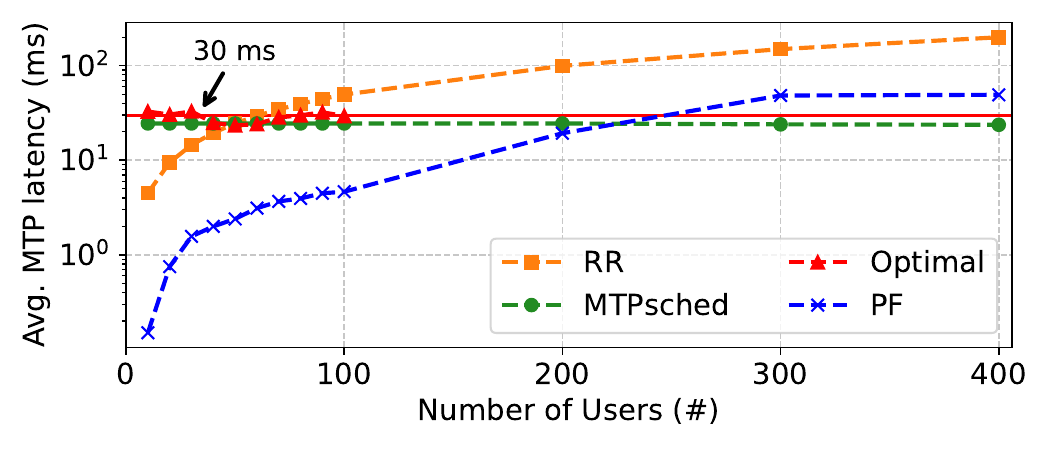}
    \caption{Average Motion-to-Photon latency as a function of the number of users.}
    \label{fig:MTP_comparing}
\end{figure}

Figure~\ref{fig:MTP_comparing} shows a comparison of the average MTP latency across all users, achieved by our MTPsched algorithm, the optimal model, and two baselines used in 5G deployments, the Round-Robin (RR) and the Proportional Fair (PF) schedule algorithms. The Round-Robin algorithm allocates resources in a cyclic order among users, without considering user's QoS requirements such as minimum throughput or latency constraints. As a result, in scenarios with a lower number of users, Round-Robin benefits from abundant resources and achieves lower average MTP latency than MTPsched and the optimal solution, which focuses on satisfying predefined latency s rather than minimizing latency. The Proportional Fair algorithm also present the similar behavior, due to the high resource availability in scenario with lower number of users.

However, as the number of users grows beyond approximately 100, the average MTP latency under the Round-Robin scheduler begins to exceed the acceptable threshold of 30 ms \cite{huawei:2018}, driven by heightened competition for limited PRBs and the lack of QoE-aware scheduling. Even the Proportional Fair algorithm starts to violate the MTP limit at around 300 users, reflecting its difficulty in handling intense resource contention. In contrast, our MTPsched consistently keeps the MTP latency below the maximum threshold, achieving near-optimal performance. This is possible because the MTPsched PRB allocation policy follows the TTI groups defined in Equation (\ref{eq:TTI_groups}). These groups are computed based on the selected frame rates, enabling MTPsched to effectively distribute resources while respecting the stringent latency requirements, even under heavy network load.

\begin{figure}
    \centering
    \includegraphics[width=1\linewidth]{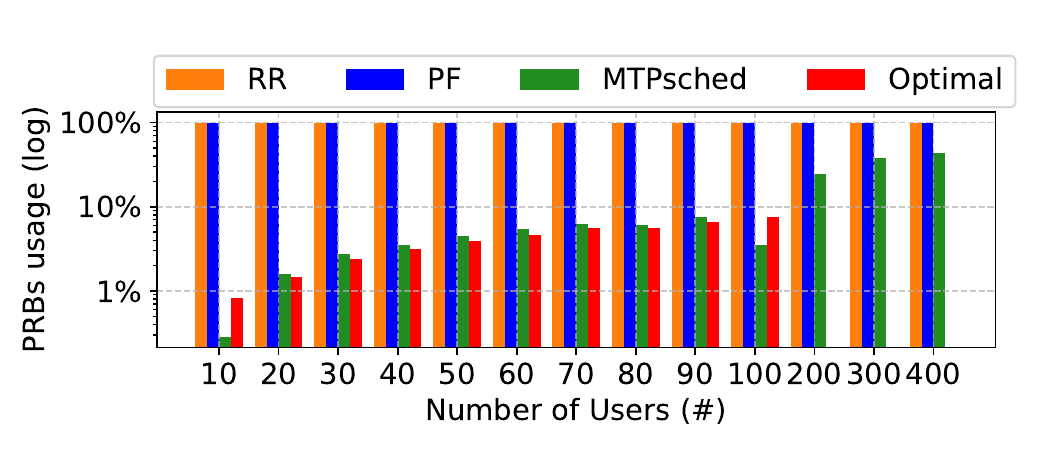}
    \caption{Normalized PRB usage for MTPsched and Round-Robin algorithms.}
    \label{fig:PRB_usage}
\end{figure}

Figure~\ref{fig:PRB_usage} presents the PRB utilization across all scenarios. Both Round-Robin and Proportional Fair schedulers allocates all PRBs available at each TTI. In contrast, both our proposed MTPsched algorithm and the optimal model strategically minimize resource usage while respecting all communication constraints. As a result, MTPsched achieves up to 75\% savings in PRB usage across most scenarios, while promoting a more balanced distribution of resources among users. Notably, in scenarios with up to 100 users, MTPsched utilizes less than 10\% of the available PRBs, while maintaining near-optimal QoE levels. Meanwhile, both Round-Robin and Proportional Fair schedulers consistently allocates all available PRBs, but still fails to meet the QoS requirements of VR-CG applications in high-density environments. It is also important to highlight that in all instances, the optimal model achieves slightly better MTP latency by consuming more PRB resources than MTPsched, achieving the best trade-off between resource efficiency and QoE satisfaction.
\section{Conclusions and Future Work}\label{sec:conclusions}

This work presented a scalable multi-stage optimization framework for resource allocation in VR-CG applications over 6G networks. We proposed three set of heuristic algorithms to efficiently solve each stage of the problem, covering user association, game engine placement, attention-aware scheduling, and MTP-aware PRB allocation. Our evaluation shows that the proposed framework achieves significant performance gains. For example, it reduces migration costs by up to 35\%, improves user QoE by over 50\%, and decreases communication resource usage by up to 75\% in comparison with baseline approaches. Moreover, our heuristics provide near-optimal solutions with an average optimality gap of 5\%, and meet the real-time computational constraints required for practical deployment in future mobile networks.

As future work, we plan to investigate the potential of AI/ML-based and metaheuristic approaches to further enhance solution adaptability. 
We are also interested in validating our proposed QoE model through Mean Opinion Score experiments, correlating objective predictions with subjective user feedback. Additionally, we recognize the relevance of combining attention-aware metrics with physiological signals, such as electrocardiogram and electroencephalogram data, to enable a more holistic and personalized assessment of QoE in immersive VR-CG scenarios.

\section*{Acknowledgments}

This work was supported by CAPES, by MCTIC/CGI. br/FAPESP under grant no. 2020/05127-2 (SAMURAI project), by CNPq under grant no. 306283/2025-5, by RNP/MCTIC under grant no. 01245.020548/2021-07 (Bra-sil 6G project), and by the OpenRAN Brazil project under grant no. A01245.014203/2021-14.



\bibliographystyle{elsarticle-num} 
\bibliography{ref}

\end{document}